\documentclass[pdflatex,sn-mathphys-num,iicol]{sn-jnl}

\usepackage{graphicx}
\usepackage{multirow}
\usepackage{amsmath,amssymb,amsfonts}
\usepackage{amsthm}
\usepackage{mathrsfs}
\usepackage[title]{appendix}
\usepackage{xcolor}
\usepackage{textcomp}
\usepackage{manyfoot}
\usepackage{booktabs}
\usepackage{algorithm}
\usepackage{algorithmicx}
\usepackage{algpseudocode}
\usepackage{listings}
\usepackage{tikz}
\usepackage{tikz-feynman}
\usepackage{physics}
\usepackage[italic]{hepnames}
\usepackage{siunitx}

\usepackage{hyperref}
\usepackage[capitalise,noabbrev]{cleveref}
\usepackage{array}
\usepackage{color}
\usepackage[usenames,dvipsnames,table]{xcolor}
\usepackage[utf8]{inputenc}
\usepackage{epsfig,relsize,xspace}
\usepackage[T1]{fontenc}
\usepackage{bbold}
\usepackage{float}
\usepackage{amsmath}
\usepackage{empheq}
\usepackage{booktabs}
\usepackage{color}
\usepackage[utf8]{inputenc}
\usepackage{xspace}
\usepackage{scalerel}
\usepackage[most]{tcolorbox}
\usepackage{multirow}
\usepackage{scalefnt}
\usepackage{bold-extra}
\usepackage[shortlabels]{enumitem}
\usepackage[tikz]{bclogo}
\usepackage{subcaption}
\usepackage{tikz-feynman}
\usepackage{cancel}
\usepackage{verbatim}
\usetikzlibrary{arrows,shapes}
\usepackage{afterpage}
\usepackage{graphicx,grffile}
\usepackage{slashed}
\usepackage{soul}
\usepackage{placeins}
\usepackage[absolute,overlay]{textpos}

\theoremstyle{thmstyleone}

\theoremstyle{thmstyletwo}

\theoremstyle{thmstylethree}

\newcommand{\beq}{\begin{equation}}
\newcommand{\eeq}{\end{equation}}

\begin{document}

\title[Article Title]{Neural simulation-based inference of the Higgs trilinear self-coupling via off-shell Higgs production}

\author[1,2,3]{\fnm{Aishik} \sur{Ghosh}}\email{AishikGhosh@physics.gatech.edu}

\author[4]{\fnm{Maximilian} \sur{Griese}}\email{maximilian.griese@mpp.mpg.de}

\author[4]{\fnm{Ulrich} \sur{Haisch}}\email{haisch@mpp.mpg.de}

\author*[4]{\fnm{Tae Hyoun} \sur{Park}}\email{taehyoun@mpp.mpg.de}

\affil[1]{\orgdiv{Department of Physics and Astronomy}, \orgname{University of California}, \orgaddress{ \city{Irvine}, \state{CA} \postcode{92697}, \country{USA}}}

\affil[2]{\orgdiv{Physics Division}, \orgname{Lawrence Berkeley National Laboratory}, \orgaddress{ \city{Berkeley}, \state{CA} \postcode{94720}, \country{USA}}}

\affil[3]{\orgdiv{School of Physics}, \orgname{Georgia Institute of Technology}, \orgaddress{ \city{Atlanta}, \state{GA} \postcode{30332}, \country{USA}}}

\affil[4]{\orgname{Max Planck Institute for Physics}, \orgaddress{\street{Boltzmannstr. 8}, \postcode{85748} \city{Garching}, \country{Germany}}}

\abstract{One of the forthcoming major challenges in particle physics is the experimental determination of the Higgs trilinear self-coupling. While efforts have largely focused on on-shell double- and single-Higgs production in proton-proton collisions, off-shell Higgs production has also been proposed as a valuable complementary probe. In this article, we design a hybrid neural simulation-based inference~(NSBI) approach to construct a likelihood of the Higgs signal incorporating modifications from the Standard Model effective field theory~(SMEFT), relevant background processes, and quantum interference effects. It leverages the training efficiency of matrix-element-enhanced techniques, which are vital for robust SMEFT applications, while also incorporating the practical advantages of classification-based methods for effective background estimates. We demonstrate that our NSBI approach achieves sensitivity close to the theoretical optimum and provide expected constraints for the high-luminosity upgrade of the Large~Hadron~Collider. While we primarily concentrate on the Higgs trilinear self-coupling, we also consider constraints on other SMEFT operators that affect off-shell Higgs production.}


\maketitle

\begin{textblock*}{25mm}[1,0](195mm,10mm)
\small MPP-2025-123
\end{textblock*}

\section{Introduction}
\label{sec:intro}

The Higgs trilinear self-coupling stands as a pivotal target in the long-term physics program of the Large Hadron Collider~(LHC). It directly probes the structure of the Higgs potential and the dynamics of electroweak symmetry breaking~(EWSB), offering a stringent test of the consistency of the Standard Model~(SM). Any precise determination --- or observed deviation from the SM prediction --- could signal the presence of new physics, such as extended scalar sectors, composite Higgs scenarios, or early-universe phenomena like electroweak~(EW) baryogenesis. A broad overview of such beyond the SM (BSM) theories can be found, for example, in the articles~\cite{DiMicco:2019ngk,Durieux:2022hbu}.

At the LHC, on-shell double-Higgs production offers a direct probe of the Higgs trilinear self-coupling, while loop-induced corrections to on-shell single-Higgs production and decay processes~\cite{Gorbahn:2016uoy,Degrassi:2016wml,Bizon:2016wgr,DiVita:2017eyz,Maltoni:2017ims,Gorbahn:2019lwq,Degrassi:2019yix,Gao:2023bll,Haisch:2024nzv} provide complementary indirect constraints. Indeed, both the ATLAS and CMS collaborations have already placed constraints on the Higgs trilinear self-coupling using inclusive measurements of these two on-shell processes, based on the complete $\sqrt{s} = 13 \, {\rm TeV}$ dataset corresponding to approximately $140 \, {\rm fb}^{-1}$ of integrated luminosity~\cite{ATLAS:2022jtk,CMS:2024awa}. Assuming that BSM effects impact only the Higgs trilinear self-coupling, the resulting limits are found to be predominantly driven by double-Higgs production, with single-Higgs processes contributing only marginally to the overall sensitivity. However, relaxing this assumption generally enhances the role of single-Higgs processes in the overall combination. 

A phenomenologically relevant example arises in BSM scenarios that simultaneously alter the Higgs trilinear self-coupling and the top-quark Yukawa coupling. In such scenarios, the double-Higgs production cross section exhibits a strong degeneracy between these two couplings, resulting in a flat direction in the two-dimensional parameter space (see, for example,~\cite{ATLAS:2022jtk,CMS:2024awa}). In contrast, single-Higgs production processes impose tight constraints on deviations of the top-quark Yukawa coupling, which can be leveraged in a combined analysis of double- and single-Higgs processes to lift this degeneracy. This example highlights the importance of incorporating as many Higgs observables as possible when constraining deviations of the Higgs trilinear self-coupling from its SM value, particularly in model-independent frameworks such as the $\kappa$-framework~\cite{LHCHiggsCrossSectionWorkingGroup:2012nn,LHCHiggsCrossSectionWorkingGroup:2013rie,LHCHiggsCrossSectionWorkingGroup:2016ypw}, the Higgs effective field theory~\cite{Feruglio:1992wf,Grinstein:2007iv,Buchalla:2012qq,Alonso:2012px,Buchalla:2013rka} or the SM effective field theory~(SMEFT)~\cite{Buchmuller:1985jz,Grzadkowski:2010es,Brivio:2017vri,Isidori:2023pyp}.

Off-shell single-Higgs production via gluon-gluon fusion~(ggF) represents an additional Higgs observable that has been shown to exhibit sensitivity to the Higgs trilinear self-coupling~\cite{Haisch:2021hvy},\footnote{The process $gg \to h^\ast \to ZZ \to 4\ell$ has also been identified as a useful indirect probe in the context of Higgs-portal models~\cite{Goncalves:2017iub,Goncalves:2018pkt,Haisch:2022rkm,Haisch:2023aiz}, as well as for testing possible modifications to light-quark Yukawa couplings~\cite{Balzani:2023jas}.} though it has not yet been experimentally utilized to constrain this coupling. The sensitivity of this channel can be greatly enhanced through the use of high-dimensional statistical inference techniques~\cite{Brehmer:2018hga,Brehmer:2018eca,Brehmer:2019xox,Cranmer:2019eaq,Stoye:2018ovl}, as demonstrated in~\cite{Ghosh_jrjc_proc} and implemented by the ATLAS collaboration~\cite{ATLAS:2024jry,ATLAS:2024ynn} to indirectly bounding the total Higgs decay width. These techniques, called neural simulation-based inference~(NSBI), require training neural networks~(NNs) on simulations and can then be used to perform statistical inference directly on unbinned and high-dimensional data. Building on~\cite{Haisch:2021hvy},\footnote{This study employed the matrix-element based kinematic discriminant from~\cite{Campbell:2013una} to enhance the sensitivity of the $pp \to ZZ \to 4 \ell$ process to potential modifications of the Higgs trilinear self-coupling.} the work at hand designs such a technique to construct an extended likelihood ratio that fully characterizes the Higgs signal in the $gg \to ZZ \to 4\ell$ channel, incorporating the leading SMEFT modifications to the Higgs trilinear self-coupling, the $q\bar{q} \to ZZ \to 4\ell$ background, and quantum interference effects between signal and background. We~demonstrate that NSBI achieves near-optimal sensitivity throughout the SMEFT parameter space considered, offering improvements over traditional histogram-based methods. Using our NSBI approach, we then assess the potential of the high-luminosity LHC~(HL-LHC) to constrain the relevant SMEFT Wilson coefficients. Although our primary emphasis is on deviations in the Higgs trilinear self-coupling, we also explore constraints on additional SMEFT operators of interest. 

This article is organized in the following way: In~\cref{sec:SMEFT}, we present the SMEFT operators relevant to the phenomenological analyses in this paper.~\cref{sec:ppZZ} gives a concise overview of the anatomy of the $pp \to ZZ$ process, along with a discussion of the Monte~Carlo~(MC) generator employed for its simulation. The~specific implementation of NSBI used in this study is described in~\cref{sec:nsbi}, while~\cref{sec:results} contains our sensitivity analysis for the HL-LHC. We~conclude and provide an outlook in~\cref{sec:conclusions}. Additional technical material is relegated to the appendix. Hey~ho,~let's~go!

\section{BSM parametrization}
\label{sec:SMEFT}

To retain model independence, we work within the SMEFT framework to capture potential BSM effects in $gg \to h^\ast \to ZZ \to 4\ell$. The effective Lagrangian we consider is given by:
\beq \label{eq:L}
{\cal L} = {\cal L}_{\rm SM} + \frac{1}{\Lambda^2} \sum_{i=H,tH,HG} C_i \hspace{0.25mm} Q_i \,.
\eeq
Here, ${\cal L}_{\rm SM}$ denotes the full SM Lagrangian, $\Lambda$~represents the common suppression scale for the dimension-six operators $Q_i$, and $C_i$ are the corresponding dimensionless Wilson coefficients, which are assumed to be real throughout this study. The~SMEFT part of the Lagrangian~(\ref{eq:L}) includes the following operators 
\beq \label{eq:operators}
\begin{split}
Q_H & = \big ( H^\dagger H \big )^3 \,, \\[0mm]
Q_{tH} & = \big ( H^\dagger H \big ) \hspace{0.5mm} \bar q \widetilde H t \,, \\[0mm]
Q_{HG} & = \big ( H^\dagger H \big ) \hspace{0.5mm} G_{\mu \nu}^a G^{a, \mu \nu} \,, 
\end{split}
\eeq
where for $Q_{tH}$ the sum of the hermitian conjugate in~(\ref{eq:L}) is understood. Here, $H$ is the Higgs doublet of the SM, and we used $\widetilde H = \varepsilon \cdot H^\ast$, with $\varepsilon = i \sigma^2$ being the antisymmetric~$SU(2)$ tensor. The symbol $q$ represents the left-handed third-generation quark $SU(2)_L$ doublets, while~$t$~denotes the right-handed top-quark $SU(2)_L$~singlet. Finally, $G_{\mu \nu}^a$ corresponds to the field strength tensor of the $SU(3)_C$ gauge group.

After EWSB the couplings relevant in the context of this article take the following form
\beq \label{eq:Lrelevant}
\begin{split}
{\cal L} & \supset -\lambda \hspace{0.25mm} \kappa_\lambda \hspace{0.25mm} v \hspace{0.075mm} h^3 - \frac{y_t}{\sqrt{2}} \hspace{0.125mm} \kappa_t \hspace{0.25mm} h \hspace{0.125mm} \bar t t \\[0mm]
& \phantom{xx} + \frac{\alpha_s}{12 \pi} \hspace{0.25mm} \kappa_g \hspace{0.25mm} \frac{h}{v} \hspace{0.5mm} G_{\mu \nu}^a G^{a, \mu \nu} \,, 
\end{split}
\eeq
where, to derive the SM contribution to the final term, we have also taken the limit of infinitely heavy top-quark mass. In~(\ref{eq:Lrelevant}), $\lambda = m_h^2/(2 v^2) \simeq 0.13$ corresponds to the value of the Higgs trilinear self-coupling in the SM, with $m_h \simeq 125 \, {\rm GeV}$ representing the Higgs mass and $v \simeq 246 \, {\rm GeV}$ the Higgs vacuum expectation value. The top-quark Yukawa coupling is given by $y_t = \sqrt{2} m_t/v$, with $m_t \simeq 166 \, {\rm GeV}$ being the top-quark mass. The symbol $\alpha_s \simeq 0.113$ represents the strong coupling constant evaluated at the Higgs boson mass scale. In terms of the Wilson~coefficients, the $\kappa$-parameters introduced in~(\ref{eq:Lrelevant}) are given by:
\beq \label{eq:kappaparameter}
\begin{split}
\kappa_\lambda & = 1 - \frac{2 \hspace{0.125mm} v^2}{m_h^2} \hspace{0.125mm} \frac{v^2}{\Lambda^2} \hspace{0.5mm} C_H \,, \\[0mm]
\kappa_t & = 1 - \frac{v}{\sqrt{2} \hspace{0.125mm} m_t} \hspace{0.125mm} \frac{v^2}{\Lambda^2} \hspace{0.5mm} C_{tH} \,, \\[0mm]
\kappa_g & = 1 + \frac{8 \pi}{\alpha_s} \hspace{0.125mm} \frac{v^2}{\Lambda^2} \hspace{0.5mm} C_{HG} \,.
\end{split}
\eeq
Here, the factors of $1$ denote the SM contributions, whereas the terms proportional to $v^2/\Lambda^2$ represent corrections arising from the SMEFT. Note that the Wilson coefficient $C_{tH}$ is normalized in terms of~$\kappa_t$ such that the top-quark mass can be used as an input parameter, as it is treated as unaffected by SMEFT contributions.

Regarding~(\ref{eq:operators}), we point out that additional SMEFT operators can, in principle, give relevant contributions to the process $gg \to h^\ast \to ZZ \to 4 \ell$. In particular, the operators $Q_{HB}$, $Q_{HW}$, and~$Q_{HW \! B}$, which modify the Higgs couplings to EW gauge bosons, contribute to this process at tree level. Notably, the ATLAS collaboration~\cite{ATLAS:2020rej,ATLAS:2025wzu} has already employed the $pp \to ZZ \to 4 \ell$ channel to place constraints on the Wilson coefficients associated with these operators. The resulting constraints are, however, substantially weaker than those derived from measurements of the $W$-boson mass, $h \to \gamma \gamma$, and $h \to \gamma Z$. For a recent discussion of this point in the context of $pp \to Z h \to \ell^+ \ell^- h$ production, see~\cite{Gauld:2023gtb}. Similar considerations hold for other potential SMEFT contributions to $gg \to h^\ast \to ZZ \to 4 \ell$, which are more effectively constrained by other measurements. Accordingly, our study focuses on the three dimension-six operators introduced in~(\ref{eq:operators}).

\section{Anatomy of $pp \to ZZ$}
\label{sec:ppZZ}

This section begins by examining the various SM and SMEFT contributions to the $pp \to ZZ$ process, and describes briefly how they are incorporated into the MC generator that provides the input data used in our NSBI analysis. We~then present numerical results for kinematic distributions --- such as the four-lepton invariant mass~$m_{4 \ell}$ --- in the process $gg \to h^\ast \to ZZ \to 4 \ell$, to illustrate the effects that arise from the three SMEFT operators defined in~(\ref{eq:operators}). 

\subsection{MC implementation}
\label{sec:MC}

\begin{figure}[!t]
\centering
\begin{subfigure}[c]{\linewidth}
\centering
\begin{tikzpicture}
\begin{feynman}
\vertex (a1) {\(g\)}; 
\vertex[right=1.5cm of a1] (a2);
\vertex[below=1.7cm of a1] (b1) {\(g\)};
\vertex[right=1.5cm of b1] (b2);
\vertex[below=0.85cm of a2] (c1);
\vertex[right=1.2cm of c1] (c2);
\vertex[right=1.1cm of c2] (h2);
\vertex[right=1.5cm of h2] (h3);
\vertex[right=1cm of h3] (h4);
\vertex[right=1cm of h4] (c3);
\vertex[right=1.0cm of c2] (f2);
\vertex[right=1.2cm of f2] (f3);
\vertex[above=0.9cm of f3] (g1) {\(Z\)};
\vertex[below=0.9cm of f3] (t1) {\(Z\)};
\diagram* {
(a1) -- [gluon, thick] (a2);
(b1) -- [gluon, thick] (b2);
(a2) -- [fermion, thick, edge label=\(t\)] (b2) -- [fermion, thick, edge label'=\(t\)] (c2) -- [fermion, thick, edge label'=\(t\)] (a2);
(c2) -- [scalar, thick, edge label=\(h\)] (f2) -- [boson, thick] (g1);
(f2) -- [boson, thick] (t1);
};
\filldraw[color=black, fill=black] (a2) circle (0.04);
\filldraw[color=black, fill=black] (b2) circle (0.04);
\filldraw[color=black, fill=black] (c2) circle (0.04);
\filldraw[color=black, fill=black] (f2) circle (0.04);
\end{feynman}
\end{tikzpicture}
\end{subfigure}

\vspace{4mm}

\begin{subfigure}[c]{\linewidth}
\centering
\begin{tikzpicture}
\begin{feynman}
\vertex (a1) {\(g\)};
\vertex[below=1.7cm of a1] (a2){\(g\)};
\vertex[right=1.5cm of a1] (a3);
\vertex[right=1.5cm of a2] (a4);
\vertex[right=1.5cm of a3] (a5);
\vertex[right=1.2cm of a5] (a7) {\(Z\)};
\vertex[right=1.5cm of a4] (a6);
\vertex[right=1.2cm of a6] (a8) {\(Z\)};
\diagram* {
{[edges=gluon]
(a1)--[thick](a3),
(a2)--[thick](a4),
},
{[edges=fermion]
(a3)--[thick, edge label=\(q\)](a5)--[thick, edge label'=\(q\)](a6)--[thick, edge label=\(q\)](a4)--[thick, edge label'=\(q\)](a3),
},
(a5) -- [boson, thick] (a7),
(a6) -- [boson, thick] (a8),
};
\filldraw[color=black, fill=black] (a3) circle (0.04);
\filldraw[color=black, fill=black] (a4) circle (0.04);
\filldraw[color=black, fill=black] (a5) circle (0.04);
\filldraw[color=black, fill=black] (a6) circle (0.04);
\end{feynman}
\end{tikzpicture}
\end{subfigure}

\vspace{4mm}

\begin{subfigure}[c]{\linewidth}
\centering
\begin{tikzpicture}
\begin{feynman}
\vertex (a1) {\(q\)};
\vertex[below=1.7cm of a1] (a2){\(\bar q\)};
\vertex[right=1.5cm of a1] (a3);
\vertex[right=1.5cm of a2] (a4);
\vertex[right=1.3cm of a3] (a5) {\(Z\)}; 
\vertex[right=1.3cm of a4] (a6) {\(Z\)};
\diagram* {
{[edges=fermion]
(a1)--[thick] (a3)--[thick, edge label'=\(q\)](a4)--[thick](a2),
},
(a3) -- [boson, thick] (a5),
(a4) -- [boson, thick] (a6),
};
\filldraw[color=black, fill=black] (a3) circle (0.04);
\filldraw[color=black, fill=black] (a4) circle (0.04);
\end{feynman}
\end{tikzpicture}
\end{subfigure}
\vspace{6mm}
\caption{\label{fig:diagramsSM} SM contributions to $pp \to ZZ$ at LO. The~top, middle, and bottom Feynman diagrams correspond to the LO processes for $gg \to h^\ast \to ZZ$, $gg \to ZZ$, and $q \bar q \to Z Z$, respectively. In the case of the first diagram, additional contributions involving bottom-quark loops are also present. All~these diagrams are included in our analysis.}
\end{figure}
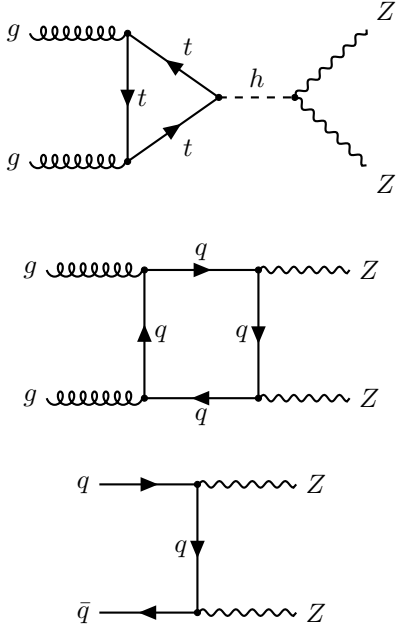

\begin{figure}[t!]
\centering
\includegraphics[width=0.45\textwidth]{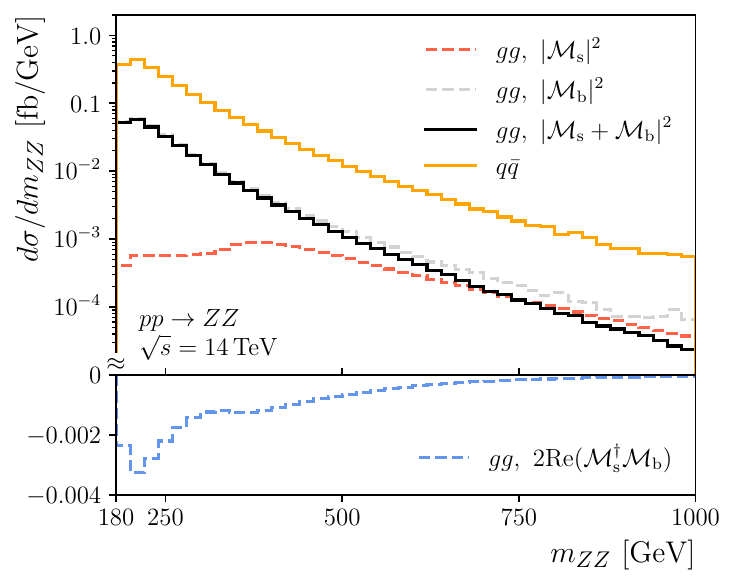}
\vspace{2mm}
\caption{The individual contributions to the~$m_{ZZ}$ distribution within the SM are shown. The signal process $gg \to h^\ast \to ZZ$~($|{\cal M}_s|^2$) is depicted by the dashed red line, while the continuum box background from $gg \to ZZ$~($|{\cal M}_b|^2$) appears as a dashed gray line. The interference term~$\big($$2 \hspace{0.25mm} {\rm Re} \left ( {\cal M}_s^\ast {\cal M}_b \right )$$\big)$ is illustrated by the blue dashed line, whereas the combined result of $gg \to (h^\ast \to) \, ZZ$ ($|{\cal M}_s + {\cal M}_b|^2$) is represented by the solid black line. Lastly, the continuum background from $q \bar q \to ZZ$ is shown as a solid orange line.}\label{fig:m4lSM} 
\end{figure}

The different leading order (LO) SM contributions to the process $pp \to ZZ$ are displayed in~\cref{fig:diagramsSM}. The~top diagram illustrates the signal process $gg \to h^\ast \to ZZ$, whereas the middle diagram depicts the continuum box background from $gg \to ZZ$. The corresponding amplitudes interfere quantum mechanically since they have the same initial and final states. Slightly above the Higgs pole, where the invariant mass of the $Z$-boson pair satisfies $m_{ZZ} \gtrsim m_h$, the interference between the signal and the continuum box background leads to a destructive contribution in the $m_{ZZ}$ distribution. This interference effect not only alters the shape of the distributions but also plays a crucial role in the signal extraction, as it enhances the sensitivity to modifications of the Higgs trilinear self-coupling in our analysis. Finally, the bottom diagram represents the tree-level continuum background $q \bar q \to ZZ$. This background reduces the purity of the Higgs signal by contributing to the same final state, but it does not interfere with the signal because it originates from a different initial~state.~\cref{fig:m4lSM} illustrates the individual SM~contributions to the $m_{ZZ}$ spectrum for proton-proton~($pp$) collisions at $\sqrt{s} = 14 \, {\rm TeV}$, assuming an integrated luminosity of $3 \, {\rm ab}^{-1}$.

\begin{figure}[!t]
\centering
\begin{subfigure}[c]{\linewidth}
\centering
\begin{tikzpicture}
\begin{feynman}
\vertex (a1) {\(g\)}; 
\vertex[right=1.5cm of a1] (a2);
\vertex[right=1.5cm of a2] (a3);
\vertex[right=1.15cm of a3] (a4);
\vertex[below=1.7cm of a1] (b1) {\(g\)};
\vertex[right=1.5cm of b1] (b2);
\vertex[right=1.5cm of b2] (b3);
\vertex[right=1.15cm of b3] (b4);
\vertex[below=0.85cm of a4] (c1);
\vertex[right=1.5cm of c1] (c2);
\vertex[right=1cm of c2] (c3);
\vertex[above=.8cm of c3] (c4){\(Z\)};
\vertex[below=.8cm of c3] (c5){\(Z\)};
\diagram* {
(a1) -- [gluon, thick] (a2);
(b1) -- [gluon, thick] (b2);
(a2) -- [fermion, edge label=\(t\), thick] (a3) -- [fermion, edge label'=\(t\), thick] (b3) -- [fermion, edge label=\(t\), thick] (b2) -- [fermion, edge label'=\(t\), thick] (a2);
(a3) -- [scalar, edge label=\(h\), thick] (c1) -- [scalar, edge label=\(h\), thick] (c2);
(b3) -- [scalar, edge label'=\(h\), thick] (c1);
(c2) -- [boson, thick] (c4);
(c2) -- [boson, thick] (c5);
};
\filldraw[color=black, fill=black] (a2) circle (0.04);
\filldraw[color=black, fill=black] (a3) circle (0.04);
\filldraw[color=black, fill=black] (b2) circle (0.04);
\filldraw[color=black, fill=black] (b3) circle (0.04);
\filldraw[color=black, fill=black] (c2) circle (0.04);
\node[square dot,fill=black,large] (d) at (c1){};
\end{feynman}
\end{tikzpicture}
\end{subfigure}

\vspace{4mm}

\begin{subfigure}[c]{\linewidth}
\centering
\begin{tikzpicture}
\begin{feynman}
\vertex (a1) {\(g\)}; 
\vertex[right=1.5cm of a1] (a2);
\vertex[below=1.7cm of a1] (b1) {\(g\)};
\vertex[right=1.5cm of b1] (b2);
\vertex[below=0.85cm of a2] (c1);
\vertex[right=1.2cm of c1] (c2);
\vertex[right=0.925cm of c2] (h2);
\vertex[right=1.2cm of h2] (h3);
\vertex[right=0.825cm of h3] (h4);
\vertex[right=1cm of h4] (c3);
\vertex[above=.8cm of c3] (c4) {\(Z\)};
\vertex[below=.8cm of c3] (c5) {\(Z\)};
 
\diagram* {
(a1) -- [gluon, thick] (a2);
(b1) -- [gluon, thick] (b2); 
(a2) -- [fermion, thick, edge label=\(t\)] (b2) -- [fermion, thick, edge label'=\(t\)] (c2) -- [fermion, thick, edge label'=\(t\)] (a2); 
(c2) -- [scalar, thick, edge label=\(h\)] (h2);
(h2) -- [scalar, half right, looseness=1.8, edge label'=\(h\), thick] (h3);
(h2) -- [scalar, half left, looseness=1.8, edge label=\(h\), thick] (h3);
(h3) -- [scalar, thick, edge label=\(h\)] (h4);
(h4) -- [boson, thick] (c4);
(h4) -- [boson, thick] (c5);
};
\filldraw[color=black, fill=black] (a2) circle (0.04);
\filldraw[color=black, fill=black] (b2) circle (0.04);
\filldraw[color=black, fill=black] (c2) circle (0.04);
\filldraw[color=black, fill=black] (h4) circle (0.04);
\node[square dot,fill=black,large] (d) at (h2){};
\node[square dot,fill=black,large] (d) at (h3){};
\end{feynman}
\end{tikzpicture}
\end{subfigure}

\vspace{4mm}

\centering
\begin{subfigure}[c]{\linewidth}
\centering
\begin{tikzpicture}
\begin{feynman}
\vertex (a1) {\(g\)}; 
\vertex[right=1.5cm of a1] (a2);
\vertex[below=1.7cm of a1] (b1) {\(g\)};
\vertex[right=1.5cm of b1] (b2);
\vertex[below=0.85cm of a2] (c1);
\vertex[right=1.2cm of c1] (c2);
\vertex[right=1.1cm of c2] (h2);
\vertex[right=1.5cm of h2] (h3);
\vertex[right=1cm of h3] (h4);
\vertex[right=1cm of h4] (c3); 
\vertex[right=1.5cm of c2] (f2);
\vertex[right=1.2cm of f2] (f3);
\vertex[above=0.9cm of f3] (g1);
\vertex[below=0.9cm of f3] (t1);
\vertex[right=1cm of g1] (c4) {\(Z\)};
\vertex[right=1cm of t1] (c5) {\(Z\)}; 
\diagram* {
(a1) -- [gluon, thick] (a2);
(b1) -- [gluon, thick] (b2);
(a2) -- [fermion, thick, edge label=\(t\)] (b2) -- [fermion, thick, edge label'=\(t\)] (c2) -- [fermion, thick, edge label'=\(t\)] (a2);
(c2) -- [scalar, thick, edge label=\(h\)] (f2) -- [scalar, edge label=\(h\), thick] (g1);
(f2) -- [scalar, edge label'=\(h\), thick] (t1) -- [boson, edge label'=\(Z\), thick] (g1);
(g1) -- [boson, thick] (c4);
(t1) -- [boson, thick] (c5);
};
\filldraw[color=black, fill=black] (a2) circle (0.04);
\filldraw[color=black, fill=black] (b2) circle (0.04);
\filldraw[color=black, fill=black] (c2) circle (0.04);
\filldraw[color=black, fill=black] (g1) circle (0.04);
\filldraw[color=black, fill=black] (t1) circle (0.04);
\node[square dot,fill=black,large] (d) at (f2){};
\end{feynman}
\end{tikzpicture}
\end{subfigure}
\vspace{6mm}
\caption{\label{fig:diagramsSMEFT2} Representative Feynman diagrams contributing at the two-loop level to the process $gg \to h^\ast \to ZZ$. The black boxes indicate insertions of the operator $Q_H$ given in~(\ref{eq:operators}).}
\end{figure}
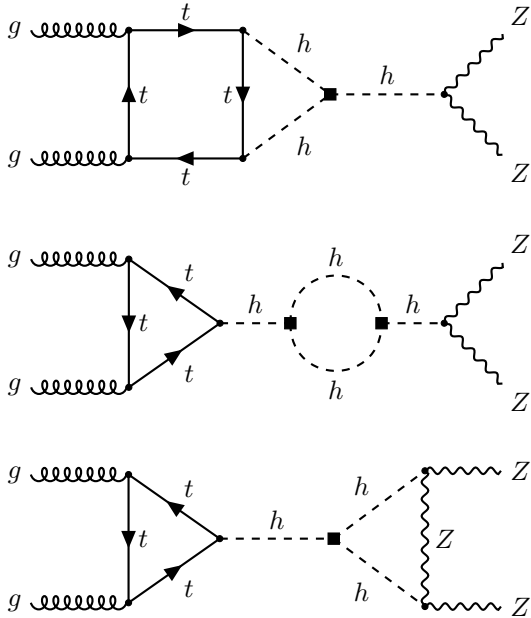

The effects stemming from insertions of the operator $Q_H$ in~(\ref{eq:operators}) are depicted in~\cref{fig:diagramsSMEFT2}. These contributions appear at the two-loop level and can be classified into three categories: first, modifications to Higgs production via ggF (top diagram); second, corrections to the Higgs propagator (middle diagram); and third, contributions affecting the Higgs decay into a pair of $Z$ bosons (bottom diagram). Compact expressions for the individual contributions have already been provided in~\cite{Haisch:2021hvy}, and are therefore not repeated here. For further details on the calculation of the individual one- and two-loop components, see also~\cite{Gorbahn:2016uoy,Degrassi:2016wml,Bizon:2016wgr,Maltoni:2017ims,Borowka:2018pxx}. We~remind the reader, however, of one important point: the contribution to the Higgs wave-function renormalization constant from propagator corrections exactly cancels with the corresponding vertex contributions when computing the full BSM correction to the off-shell $gg \to h^\ast \to ZZ$ amplitude. This cancellation is expected, as the Higgs appears only as an internal particle in this process. Consequently, the only two-loop term quadratic in the Wilson coefficient $C_H$ originates from the bare Higgs self-energy being the sole momentum-dependent correction of this kind. This contrasts with on-shell Higgs processes, where each external Higgs leg is accompanied by a corresponding wave-function renormalization factor.

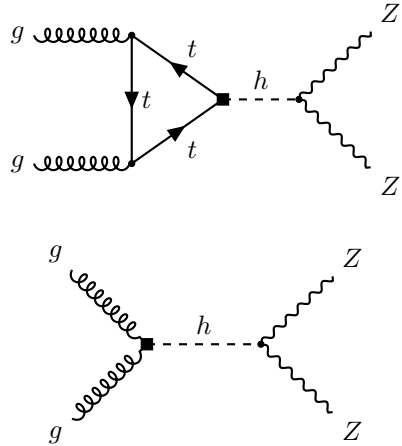
\begin{figure}[!t]
\centering
\begin{subfigure}[c]{\linewidth}
\centering
\begin{tikzpicture}
\begin{feynman}
\vertex (a1) {\(g\)}; 
\vertex[right=1.5cm of a1] (a2);
\vertex[below=1.7cm of a1] (b1) {\(g\)};
\vertex[right=1.5cm of b1] (b2);
\vertex[below=0.85cm of a2] (c1);
\vertex[right=1.2cm of c1] (c2);
\vertex[right=1.1cm of c2] (h2);
\vertex[right=1.5cm of h2] (h3);
\vertex[right=1cm of h3] (h4);
\vertex[right=1cm of h4] (c3); 
\vertex[right=1.0cm of c2] (f2);
\vertex[right=1.2cm of f2] (f3);
\vertex[above=0.9cm of f3] (g1) {\(Z\)};
\vertex[below=0.9cm of f3] (t1) {\(Z\)}; 
\diagram* {
(a1) -- [gluon, thick] (a2);
(b1) -- [gluon, thick] (b2);
(a2) -- [fermion, thick, edge label=\(t\)] (b2) -- [fermion, thick, edge label'=\(t\)] (c2) -- [fermion, thick, edge label'=\(t\)] (a2);
(c2) -- [scalar, thick, edge label=\(h\)] (f2) -- [boson, thick] (g1);
(f2) -- [boson, thick] (t1);
};
\filldraw[color=black, fill=black] (a2) circle (0.04);
\filldraw[color=black, fill=black] (b2) circle (0.04);
\filldraw[color=black, fill=black] (f2) circle (0.04);
\node[square dot,fill=black,large] (d) at (c2){};
\end{feynman}
\end{tikzpicture}
\end{subfigure}

\vspace{4mm}

\begin{subfigure}[c]{\linewidth}
\centering
\begin{tikzpicture}
\begin{feynman}
\vertex (a1) {\(g\)}; 
\vertex[below=2.4cm of a1] (b1) {\(g\)};
\vertex[below=1.2cm of a2] (c1);
\vertex[right=-0.3cm of c1] (c2);
\vertex[right=1.4cm of c2] (h2); 
\vertex[right=1.5cm of c2] (f2);
\vertex[right=1.2cm of f2] (f3);
\vertex[above=0.9cm of f3] (g1) {\(Z\)};
\vertex[below=0.9cm of f3] (t1) {\(Z\)}; 
\diagram* {
(a1) -- [gluon, thick] (c2);
(b1) -- [gluon, thick] (c2);
(c2) -- [scalar, thick, edge label=\(h\)] (f2) -- [boson, thick] (g1);
(f2) -- [boson, thick] (t1);
};
\filldraw[color=black, fill=black] (f2) circle (0.04);
\node[square dot,fill=black,large] (d) at (c2){}; 
\end{feynman}
\end{tikzpicture}
\end{subfigure}
\vspace{4mm}
\caption{\label{fig:diagramsSMEFT1} Illustrative Feynman diagrams for $gg \to h^\ast \to ZZ$, showing contributions from the operators $Q_{tH}$ (top panel) and $Q_{HG}$ (bottom panel). Operator insertions are marked with black boxes. }
\end{figure}

The contributions arising from insertions of the operators $Q_{tH}$ and $Q_{HG}$ in~(\ref{eq:operators}) are shown in~\cref{fig:diagramsSMEFT1}. These effects enter at the one-loop and tree level, respectively. The corresponding amplitudes can be directly derived from the top-quark contribution to $gg \to h^\ast \to ZZ$ within the SM and the associated expression in the limit of infinitely heavy top-quark mass. This conclusion is a direct consequence of~(\ref{eq:Lrelevant}) and~(\ref{eq:kappaparameter}).

In order to predict $pp \to ZZ$, we have modified the MC generator {\tt MCFM~10.3}~\cite{Boughezal:2016wmq}. Out of the box, this code includes all the SM amplitudes described above and shown in~\cref{fig:diagramsSM}. The SM corrections are augmented by the scattering amplitudes corresponding to the contributions illustrated in~Figures~\ref{fig:diagramsSMEFT2} and~\ref{fig:diagramsSMEFT1}. The modified code is thus capable of computing arbitrary differential distributions for the full $pp \to ZZ$ process, including corrections arising from the Wilson coefficients~$C_H$, $C_{tH}$, and $C_{HG}$. We have also implemented an interface that enables the computation of the various squared matrix elements~(MEs) used as input for our NSBI analysis. The corresponding modified {\tt MCFM~10.3} source code is publicly accessible at~\cite{ourgit}. Our new BSM implementation of the effects involving the Wilson coefficient $C_H$ has been successfully validated against the original {\tt MCFM~8.0} implementation presented in~\cite{Haisch:2021hvy}, as well as the {\tt MCFM~7.0} implementation integrated within the {\tt JHUGen~MELA} framework~\cite{Gritsan:2020pib}.

\subsection{Kinematic distributions}
\label{sec:distributions}

Throughout our analysis, we adopt the following input parameters: $G_F = 1/(\sqrt{2} \hspace{0.25mm} v^2) = 1.166379 \cdot 10^{-5} \, {\rm GeV}^{-2}$, $m_W = 80.37 \, {\rm GeV}$, $m_Z = 91.188 \, {\rm GeV}$, and $m_h = 125.2\, {\rm GeV}$~\cite{ParticleDataGroup:2024cfk}. The electromagnetic coupling constant $\alpha$ and the square of the sine of the weak mixing angle $\sin^2 \theta_w$ are computed using the $G_F$ scheme~\cite{Denner:2000bj}. For~the top and bottom quark masses, we use $m_t = 166 \, {\rm GeV}$, and $m_b = 2.8 \, {\rm GeV}$, corresponding to $\overline{\rm MS}$ masses evaluated at the Higgs mass. The presented spectra correspond to $pp$ collisions at $\sqrt{s} = 14 \, {\rm TeV}$ with an integrated luminosity of $3 \, {\rm ab}^{-1}$, using the {\tt NNPDF40\_nlo\_as\_01180} parton distribution functions~(PDFs)~\cite{NNPDF:2021njg}. The renormalization and factorization scales are dynamically set to~$m_{ZZ}$, on an event-by-event basis. Our predictions account for both different-~($e^+ e^- \mu^+ \mu^-$) and same-flavor~($2 e^+ 2 e^-$ and $2 \mu^+ 2 \mu^-$) final states from the decays of the two $Z$ bosons. 

To include next-to-leading order (NLO) QCD corrections in our ggF prediction, we adopt the findings of~\cite{Buonocore:2021fnj}, which demonstrated that the ratio of NLO to LO predictions for $gg \to ZZ \to 4 \ell$ is essentially flat in $m_{ZZ}$. By averaging over the ratio of the NLO and LO $m_{ZZ}$ spectra, one obtains $K_{gg}^{\rm NLO} = 1.83$. Assuming that QCD and EW interactions factorize --- which is expected to be an accurate approximation in this context --- we use this factor to normalize the distributions in both the SM and SMEFT scenarios. In the case of the $q \bar q \to ZZ$ channel, we employ in our analysis the ratio of the next-to-next-to-leading order~(NNLO) QCD to LO cross sections from~\cite{Grazzini:2018owa}. The ratio is $K_{q \bar q}^{\rm NNLO} = 1.55$ and found to be nearly constant across the $m_{ZZ}$ spectrum.

We focus on the off-shell region by selecting events within $m_{ZZ} > 180 \, {\rm GeV}$. In this context, $m_{ZZ}$ serves as a proxy for the four-lepton invariant mass~$m_{4 \ell}$. Leptons ($\ell = e, \mu$) are required to lie within the pseudorapidity range $|\eta_\ell| < 2.5$. The lepton with the highest transverse momentum ($p_T$) must satisfy $p_T^{\ell_1} > 20 \, {\rm GeV}$, while the second, third, and fourth leptons in $p_T$ order must satisfy $p_T^{\ell_2} > 15 \, {\rm GeV}$, $p_T^{\ell_3} > 10 \, {\rm GeV}$, and $p_T^{\ell_4} > 7 \, {\rm GeV}$, respectively. The dilepton pair whose invariant mass is closest to the $Z$-boson mass is designated as the leading dilepton pair, and its mass is required to lie within the range $50 \, {\rm GeV} < m_{12} < 106 \, {\rm GeV}$. The subleading dilepton pair must satisfy $50 \, {\rm GeV} < m_{34} < 115 \, {\rm GeV}$. These selection criteria are similar to those used in the latest ATLAS and CMS analyses of off-shell Higgs production in~ggF~\cite{ATLAS:2024jry,CMS:2024eka}. 

\begin{figure}[t!]
\centering
\includegraphics[width=0.45\textwidth]{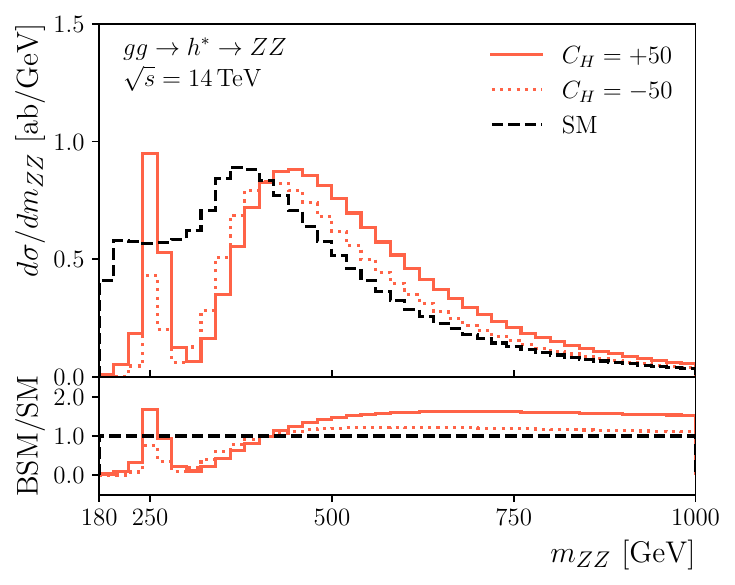}
\includegraphics[width=0.45\textwidth]{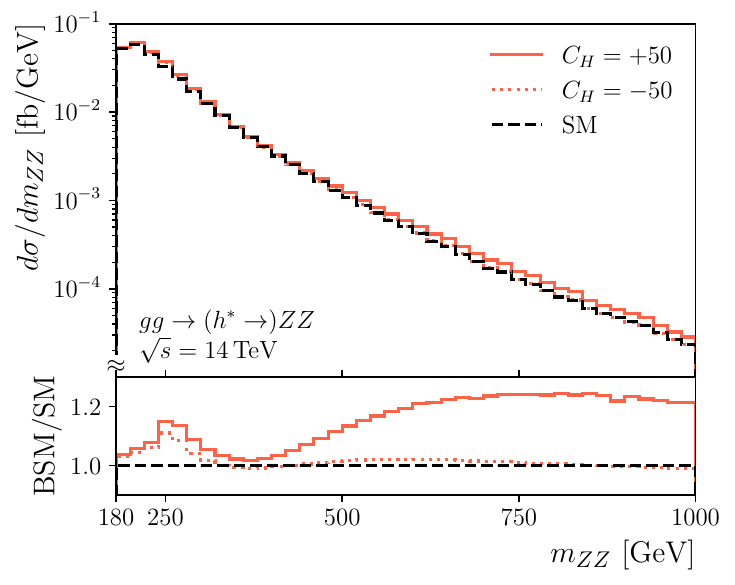}
\vspace{4mm}
\caption{\label{fig:m4lCH} Upper panel: Distributions of $m_{ZZ}$ for the Higgs signal in the SM~(dashed black) and with BSM corrections for $C_H = +50$ (solid orange) and $C_H = -50$ (dotted orange). Lower panel: Same as the upper panel, but including the Higgs signal, the continuum box background, and their interference. The ratio plots beneath each panel show the distributions normalized to the corresponding SM prediction. All results are obtained under the assumption $C_{tH} = C_{HG} = 0$, with a common suppression scale of $\Lambda = 1 \, {\rm TeV}$ for the dimension-six operators.}
\end{figure}

\cref{fig:m4lCH} displays $m_{ZZ}$ distributions for the Higgs channel alone in the upper panel, and for the combined contributions of the Higgs channel, the continuum box background, and their interference in the lower panel. The displayed BSM predictions correspond to two different values of~$C_H$, with $C_{tH} = C_{HG} = 0$ and $\Lambda = 1 \, {\rm TeV}$. A prominent feature in the upper panel is that for $275 \, {\rm GeV} \lesssim m_{ZZ} \lesssim 400 \, {\rm GeV}$, the BSM distributions for $gg \to h^\ast \to ZZ \to 4 \ell$ are noticeably smaller than the SM prediction. This behavior can be attributed to the fact that, for sufficiently large values of the Wilson coefficient~$C_H$, the dominant BSM contributions arise from corrections to the Higgs propagator --- these are the only corrections quadratic in $C_H$. Importantly, the propagator corrections necessarily reduce the real part of the $gg \to h^\ast \to ZZ \to 4 \ell$ amplitude within the relevant $m_{ZZ}$ range. This leads to a destructive interference that can become so significant that the BSM contribution nearly cancels the SM amplitude. Above the two-Higgs production threshold at $m_{ZZ} = 2 m_h$, the Higgs self-energy acquires an imaginary part, as both internal Higgs lines in the propagator correction can go on-shell. For the large values of $C_H$ shown in the figure, this results in a resonance-like feature in the $m_{ZZ}$ distribution near the two-Higgs production threshold. When propagator corrections are dominant, they shift the peak of the distribution toward higher $m_{ZZ}$ values and enhance the high-mass tail of the spectrum. This feature is clearly visible for $C_H = +50$, but much less so for $C_H = -50$, where additional BSM contributions from Higgs production and decay also play a role, diluting the overall~effect.

The lower panel of~\cref{fig:m4lCH} presents our results for the $m_{ZZ}$ distributions in the $gg \to (h^\ast \to) \, ZZ \to 4 \ell$ process. A notable feature in both BSM spectra is the peak-like structure around the two-Higgs production threshold, which arises from the interference between the signal and the background. This unusual shape deformation once again arises from corrections to the Higgs propagator. As shown in the articles~\cite{Haisch:2022rkm,Haisch:2023aiz}, such distortions offer a unique handle on loop-level contributions to the Higgs self-energy, especially those involving light virtual particles, as found for instance in Higgs-portal scenarios. Depending on the specific value of~$C_H$, the $m_{ZZ}$ distribution may also exhibit an enhancement in its high-mass tail. This behavior is clearly illustrated in the plot for the case $C_H = +50$. It is important to note that the BSM features observed here differ qualitatively from the modifications induced by tree-level insertions of dimension-six SMEFT operators $\big($see,~for example,~\cite{Gainer:2014hha,Englert:2014ffa}$\big)$, which typically exhibit an approximately quadratic growth with~$m_{ZZ}$. The~same feature is also exploited in indirect determinations of the total Higgs decay width, as in, for instance,~\cite{ATLAS:2024jry,CMS:2024eka}.

\begin{figure}[t!]
\centering
\includegraphics[width=0.45\textwidth]{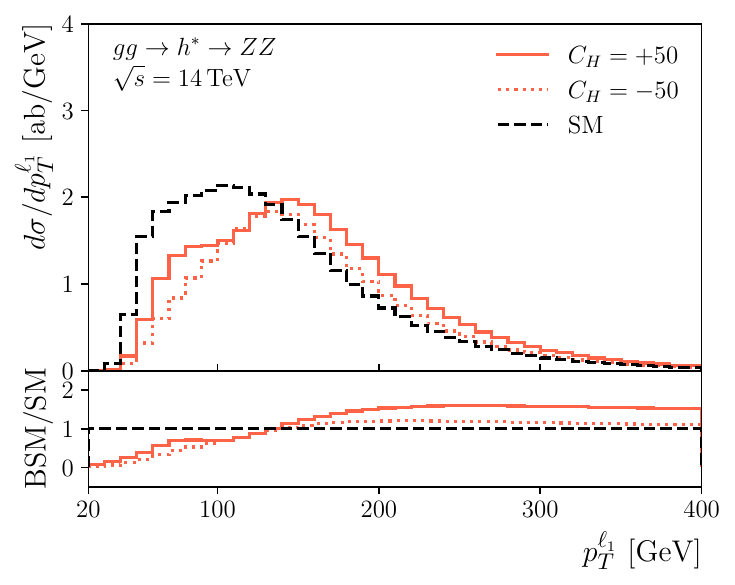}
\includegraphics[width=0.45\textwidth]{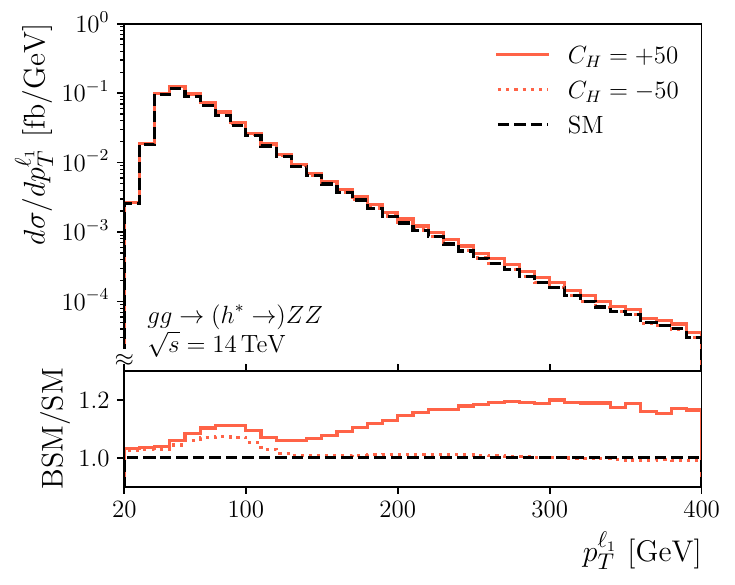}
\vspace{4mm}
\caption{\label{fig:pTl1CH} Same as~\cref{fig:m4lCH}, but showing the transverse momentum of the leading lepton $p_T^{\ell_1}$.}
\end{figure}

Beyond its impact on the $m_{ZZ}$ distribution, the operator $Q_H$ also influences other kinematic observables in off-shell Higgs production through~ggF. To illustrate this,~\cref{fig:pTl1CH} presents the $p_T^{\ell_1}$ spectra for both the SM and two BSM scenarios with a non-zero Wilson coefficient $C_H$. As before, the upper panel shows the signal process $gg \to h^\ast \to ZZ \to 4\ell$, while the lower panel presents the complete $gg \to (h^\ast \to) \, ZZ \to 4\ell$ results, including the signal, background, and their interference. For $C_H$, the distortions in the~$p_T^{\ell _1}$ spectrum resemble the changes seen in the~$m_{ZZ}$ distributions shown in~\cref{fig:m4lCH}. However, since $p_T^{\ell_1}$ unlike $m_{ZZ}$, does not directly probe the two-Higgs production threshold, the resonance-like structure is less pronounced at low~$p_T^{\ell_1}$. Similar patterns are also observed in the~$p_T$ distributions of the remaining leptons. This implies that, for the operator $Q_H$, the modifications in the lepton $p_T$ spectra and the~$m_{ZZ}$ distributions are driven by essentially the same underlying physics.

In~Figures~\ref{fig:m4lCH} and~\ref{fig:pTl1CH}, we have shown the $m_{ZZ}$ and~$p_T^{\ell_1}$ distributions for the two benchmark choices $C_H = \pm 50$. These values are selected simply because they induce pronounced distortions in the spectra that can be clearly distinguished by eye from the SM expectations. In~Appendix~\ref{sec:additionalc6Benchmarks}, we perform the same study for $C_H = \pm 10$, highlighting how the shape of the BSM contributions to the distributions changes as the absolute value of the Wilson coefficient $C_H$ decreases.

\begin{figure}[t!]
\centering
\includegraphics[width=0.49\textwidth]{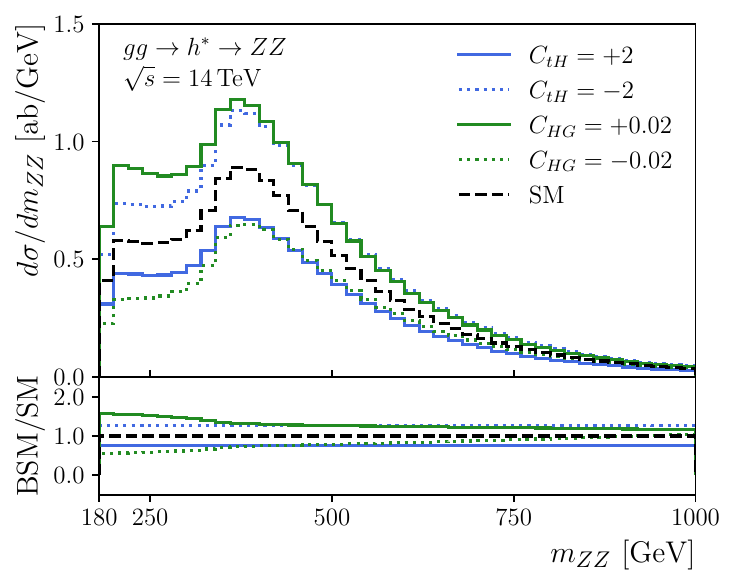}
\includegraphics[width=0.49\textwidth]{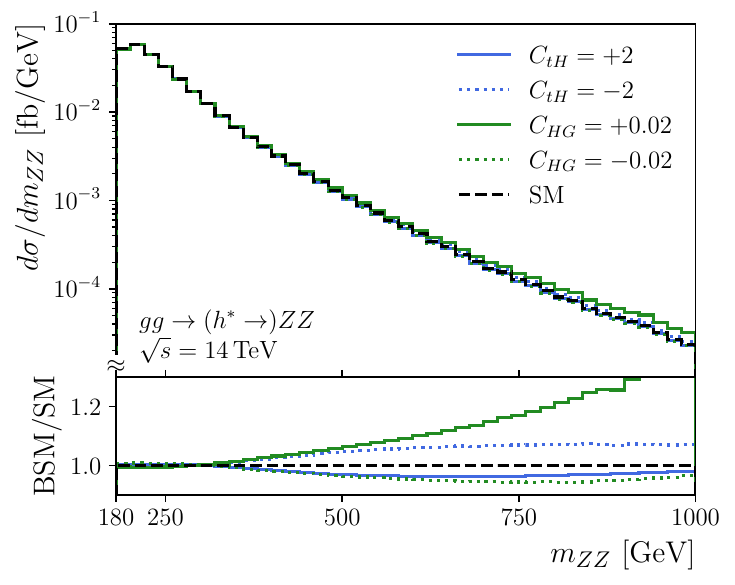}
\vspace{4mm}
\caption{\label{fig:m4lCtHCHG} Upper panel: $m_{ZZ}$ spectra for the Higgs signal in the SM (dashed black) and with BSM effects for the choices $C_{tH} = +2$~(solid blue), $C_{tH} = -2$~(dotted blue), $C_{HG} = +0.02$ (solid green), and $C_{HG} = -0.02$ (dotted green). Lower panel: Same as above, but including the continuum background and interference. The ratio plots show distributions normalized to the SM. All Wilson coefficients not explicitly specified are set to zero, and the presented results assume $\Lambda = 1 \, {\rm TeV}$.}
\end{figure}

\begin{figure}[t!]
\centering
\includegraphics[width=0.49\textwidth]{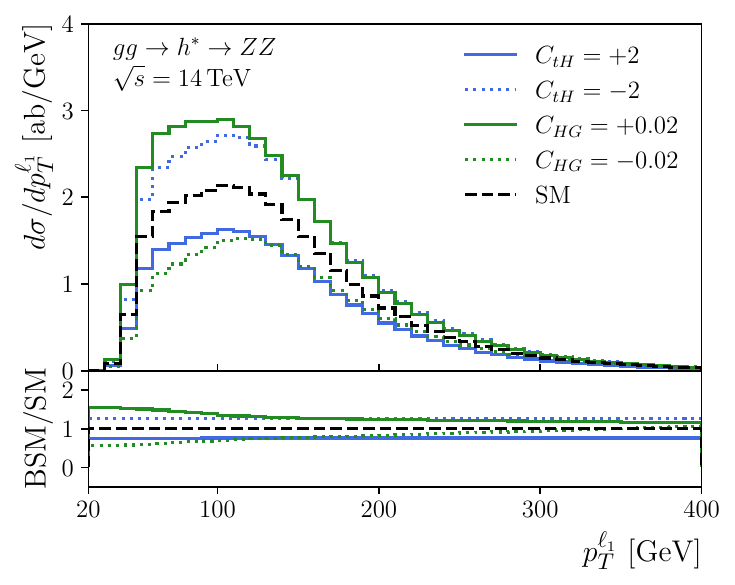}
\includegraphics[width=0.49\textwidth]{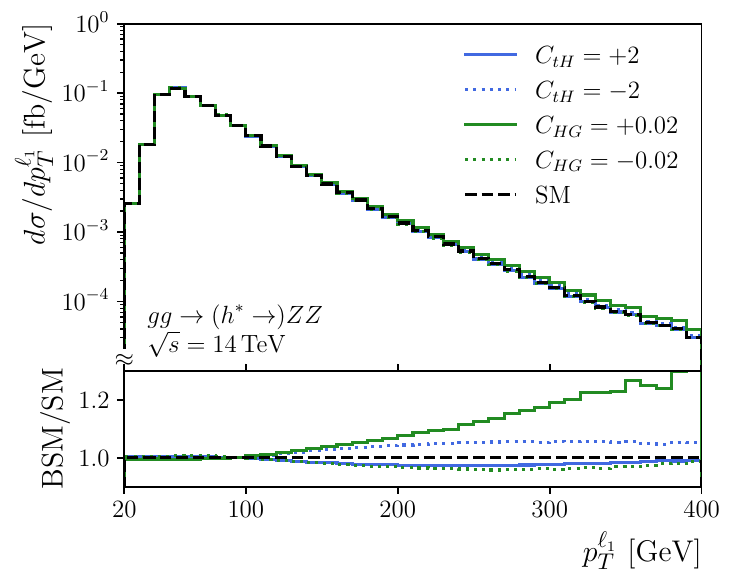}
\vspace{4mm}
\caption{\label{fig:pTl1CtHCHG} Identical to~\cref{fig:m4lCtHCHG}, but displaying the transverse momentum of the leading lepton $p_T^{\ell_1}$.}
\end{figure}

Figures~\ref{fig:m4lCtHCHG} and~\ref{fig:pTl1CtHCHG} compare the effects of non-zero~$C_{tH}$ and $C_{HG}$ on the $m_{ZZ}$ and $p_T^{\ell_1}$ distributions. The upper panel of~\cref{fig:m4lCtHCHG} shows that~$C_{tH}$ effectively rescales the $m_{ZZ}$ distribution for the process $gg \to h^\ast \to ZZ \to 4 \ell$ compared to the SM prediction. This behavior reflects the fact that the relevant amplitude is directly proportional to the top-quark contribution in the SM. With our choice of $C_{tH} = +2$ ($C_{tH} = -2$), this leads to an overall suppression (enhancement) of the differential rate by a factor of 0.75 (1.25). The~modifications due to $C_{HG}$ are instead not flat in~$m_{ZZ}$. This is because the $gg \to h^\ast \to ZZ \to 4\ell$ amplitude induced by an insertion of~$Q_{HG}$ is independent of the top-quark mass, unlike the SM amplitude, which depends on it. Specifically, the SM amplitude is enhanced near the top-quark pair production threshold, whereas the BSM contribution shows no such enhancement. This accounts for the fact that, for $C_{HG} = +0.02$ ($C_{HG} = -0.02$), the ratio of BSM to SM distributions decreases (increases) from approximately 1.5 (0.5) at $m_{ZZ} \simeq 200 \, {\rm GeV}$ to about~1.25~(0.75) at $m_{ZZ} \simeq 2 m_t$.

The~lower panel of~\cref{fig:m4lCtHCHG} illustrates that the effects of~$Q_{tH}$ and~$Q_{HG}$ also modify the $m_{ZZ}$ distribution of $gg \to (h^\ast \to) \, ZZ \to 4\ell$ differently. The~primary differences appear in the high-$m_{ZZ}$ tail of the spectrum. For example, with $C_{tH} = -2$ and $C_{HG} = +0.02$, the SMEFT predictions coincide at $m_{ZZ} \simeq 500 \, {\rm GeV}$. However, at higher $m_{ZZ}$ values, the differential cross section for $C_{HG} = +0.02$ surpasses that for $C_{tH} = -2$, with the discrepancy increasing quadratically with $m_{ZZ}$. This difference arises from the form factor suppression of the~$Q_{tH}$ contribution, connected to the substructure of the gluon-gluon-Higgs three-point function involving virtual top quarks. Since the operator $Q_{HG}$ is defined as point-like in the Lagrangian~(\ref{eq:operators}), it does not experience this type of suppression. It is inherently treated as a contact interaction, which cannot be further resolved by the produced Higgs boson or by additional QCD radiation. This effect is amplified by the interference between the signal and the continuum box background contributing to the process $gg \to (h^\ast \to) \, ZZ \to 4\ell$. 

The impact of the operators~$Q_{tH}$ and~$Q_{HG}$ on the $p_T^{\ell_1}$ spectrum are depicted in~\cref{fig:pTl1CtHCHG}. Both panels show that the structure of BSM effects relative to the SM in the $p_{T}^{\ell_1}$ distribution closely parallels the behavior seen in the corresponding $m_{ZZ}$ spectra. As in~\cref{fig:m4lCtHCHG}, the most prominent feature is the quadratic growth of the result for $C_{HG} = +0.02$ present in the results for $gg \to (h^\ast \to) \, ZZ \to 4\ell$. This behavior again stems from the point-like nature of the $Q_{HG}$ operator, which cannot be resolved by quantum fluctuations, and is enhanced by its interference with the continuum box contribution.

We also examined the distributions of all other observables characterizing the process $pp \to ZZ \to 4\ell$ $\big($a comprehensive list can be found, for example, in~\cite{ATLAS:2024jry}$\big)$, but observed that they are generally less sensitive to modifications induced by the three dimension-six SMEFT operators introduced in~(\ref{eq:operators}). Since our NSBI approach employs the squared MEs, which encapsulate the full kinematic information of each event, we will not present plots for other kinematic observables, such as the pseudorapidities or the angles of the final-state leptons in the azimuthal plane.

\section{NSBI analysis}
\label{sec:nsbi}

In this section, we present the main elements of our NSBI analysis applied to the processes $pp \to ZZ \to 4\ell$ and $pp \to ZZ \to 2\ell 2\nu$. NSBI~uses NNs that analyze multiple kinematic observables to estimate the probability density ratio of individual events under different hypotheses for statistical testing. As we will explain, carefully choosing the targets for the NNs within the probability mixture model and enhancing prediction accuracy through calibration are essential for ensuring the reliability of the inference process.

\subsection{Likelihood ratio estimation}
\label{sec:likelihoodratio}

The likelihood of a parameter $\theta$, describing the total rate and distribution of events characterized by a set of experimental observables $x$, is given by:
\beq \label{eq:likelihood}
{\cal L}\left (\theta | x \right ) = {\cal P} \left (n ; \nu(\theta) \right ) \prod_{i = 1}^{n} p \left (x_i | \theta \right ) \,.
\eeq
Here, ${\cal P} \left(n ; \nu(\theta) \right) = \nu^n(\theta) \exp \left(-\nu(\theta)\right)/n!$ denotes the Poisson probability of observing $n$ events in the dataset, given an expected number $\nu(\theta)$ under the hypothesis. According to the Neyman-Pearson Lemma, the most powerful test statistic for distinguishing between the hypotheses $\theta_1$ and $\theta_2$ is the ratio of their likelihoods:
\beq \label{eq:likeratio}
\lambda \left (\theta_1, \theta_2 \right ) = \frac{{\cal L} \left (\theta_1 | x \right )}{{\cal L} \left (\theta_2 | x \right )}.
\eeq 

In particle physics, the connection between the ME ${\cal M}(z; \theta)$ of a process and its corresponding probability distribution~$p \left (x | \theta \right )$ is expressed as:
\beq \label{eq:highenergy}
\frac{d\sigma}{dx} = \int \! dz \, p \left (x|z \right ) \left | {\cal M} \left (z ; \theta \right ) \right |^2 = \sigma(\theta) \hspace{0.5mm} p \left (x | \theta \right ) \,.
\eeq
In this context, the product of the total cross section $\sigma(\theta)$ and the integrated luminosity $L$ collected by the collider experiment determines the total expected number of events, given by $\nu(\theta) = \sigma(\theta) \, L$. This may include acceptance and efficiency effects related to the detector and the analysis. The joint probability density $p(x, z | \theta) = p(x | z) \left| {\cal M}(z; \theta) \right|^2 / \sigma(\theta)$ captures the evolution of the latent variables $z$ through a convolution that incorporates complex processes such as PDFs, parton showering, hadronization, and detector effects. Although the ME of a process predicts both the rate (via the Poisson term) and shape (through the product of probability densities) components of the likelihood, expressing the latter directly in terms of experimental observables is numerically intractable. The NSBI technique makes these likelihood ratios tractable, enabling hypothesis tests to be carried out with optimal statistical power.

\subsection{Probability mixture model}
\label{sec:probmix}

\begin{figure*}[!t]
\centering
\includegraphics[width=\textwidth,height=0.5\textheight,keepaspectratio]{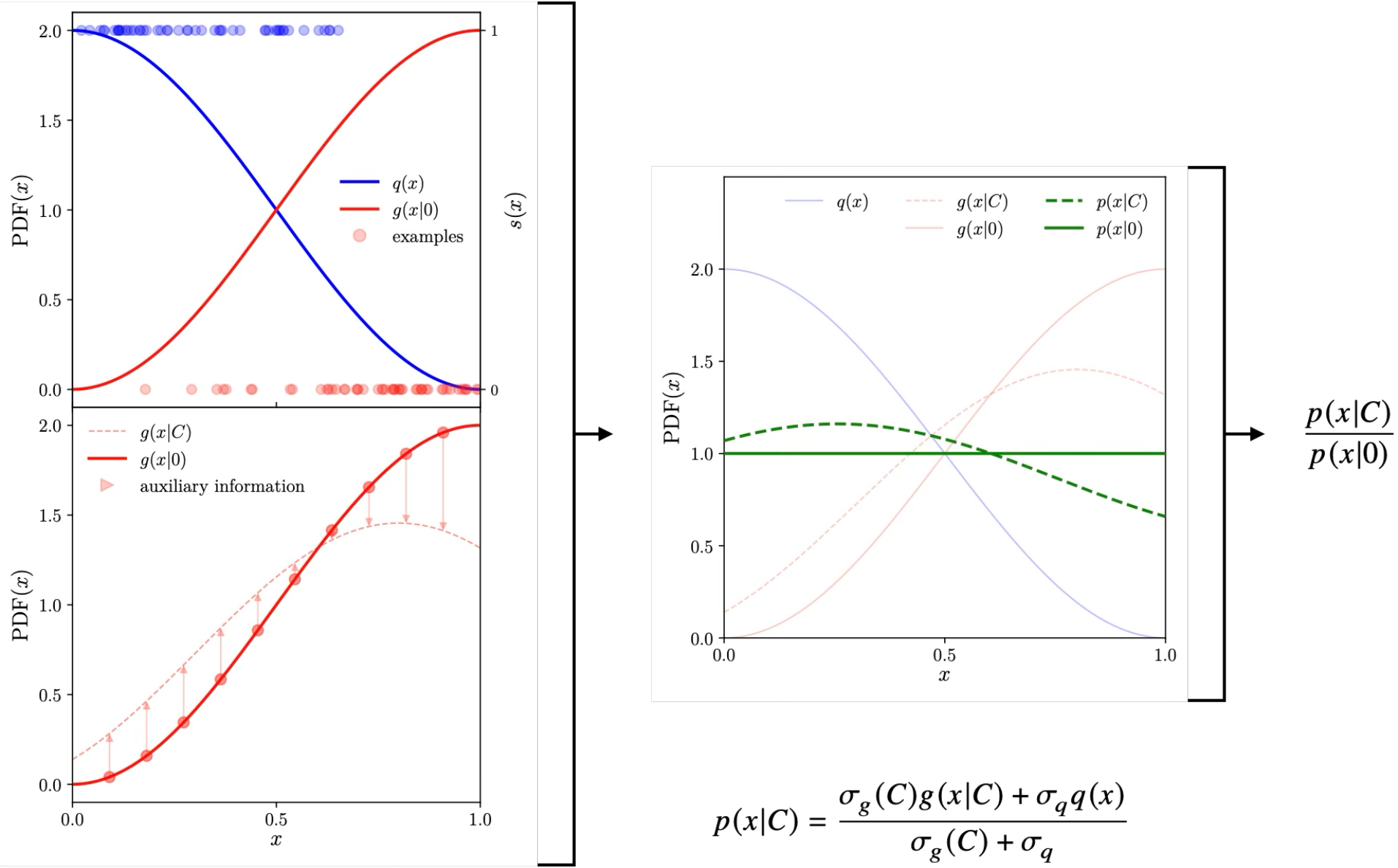}
\vspace{4mm}
\caption{Conceptual diagram of the mixture model employed in this work, combining two probability ratio estimates~(left). The first is trained to distinguish between samples from discrete hypotheses~(top), while the second learns a continuous hypothesis space using auxiliary information from the MC generator~(bottom). Their combination through the mixture model (middle) provides an estimate of the inclusive probability ratio~(right). The presented probability density functions~(PDFs) are illustrative examples.} \label{fig:mixture_model}
\end{figure*}

The probability of a collider process involving multiple partonic initial-state configurations $i$ can be represented by the following mixture model:
\beq \label{eq:prob_mixture_model}
p \left (x | \theta \right ) = \frac{ \sum_i \sigma_i(\theta) \hspace{0.5mm} p_i \left (x | \theta \right )}{\sum_i \sigma_i (\theta)} \,.
\eeq
For the $pp \to ZZ \to 4 \ell$ and $pp \to ZZ \to 2 \ell 2\nu$ processes, the mixture comprises of two distinct contributions: $gg$-initiated corrections, which depend on the Wilson coefficient set~$C = \big\{C_H, C_{tG}, C_{HG}\big\}$, and $q \bar q$-initiated corrections, which are independent of $C$. Accordingly, the appropriate mixture model takes the form:
\beq \label{eq:prob_pp_mixture_model}
p \left (x | C \right ) = \frac{\sigma_{g}(C) \hspace{0.5mm} g \left (x | C \right ) + \sigma_{q} \hspace{0.5mm} q \left (x \right ) }{\sigma_{g}(C)+ \sigma_{q}} \,.
\eeq
The first term in~(\ref{eq:prob_pp_mixture_model}) incorporates all quantum interference effects discussed in~\cref{sec:ppZZ}. Separate NNs are trained to estimate each term in this model, as detailed in the following subsection.

To approximate the ratio between probabilities of an event under two different hypotheses, a binary classifier can be trained using samples from the hypotheses with equal priors, i.e. satisfying $\nu(\theta_1) = \nu(\theta_2)$. In the asymptotic limit, the classifier decision can be transformed into the ratio of the two probability densities:
\beq \label{eq:carl}
r \left (x; \theta_1, \theta_2 \right ) = \frac{p \left (x | \theta_1 \right )}{ p \left (x | \theta_2 \right ) } \approx \frac{s \left (x \right )}{1 - s \left (x \right )} \,.
\eeq
Here, $s \left (x \right )$ denotes the classifier decision function~\cite{Sugiyama_Suzuki_Kanamori_2012,cranmer2016approximatinglikelihoodratioscalibrated}. For numerical stability and broad coverage of the feature space, it is convenient to select a common denominator hypothesis whose support overlaps significantly with that of the numerator. In the context of this study, the SM $gg \to (h^\ast \to) \, ZZ$ process is used as the denominator hypothesis when estimating probability ratios for the $q\bar{q} \to ZZ$ and $q\bar{q} \to W^{+}W^{-}$ processes.

To perform inference over the continuous parameter space spanned by the Wilson coefficient set $C$, we define a rate-normalized probability ratio relative the SM (associated with the hypothesis~$\theta_0$) as:
\beq \label{eq:fratio}
R \left (x ; \theta - \theta_0 \right ) = \frac{\sigma(\theta)}{\sigma(\theta_0)} \hspace{0.25mm} \frac{p \left (x | \theta \right )}{p \left (x | \theta_0 \right )} \,.
\eeq 
This function is then expanded as a Taylor series around $\theta_0$ to fourth order in $\delta = \theta - \theta_0$:
\begin{eqnarray} \label{eq:taylor}
\begin{split}
R \left (x ; \delta \right ) & \approx a_0(x) + a_i(x) \hspace{0.25mm} \delta_i + a_{ij}(x) \hspace{0.25mm} \delta_i \hspace{0.25mm} \delta_j \\[2mm]
& \phantom{xx} + a_{ijk}(x) \hspace{0.25mm} \delta_i \hspace{0.25mm} \delta_j \hspace{0.25mm} \delta_k + a_{ijkl}(x) \hspace{0.25mm} \delta_i \hspace{0.25mm} \delta_j \hspace{0.25mm} \delta_k \hspace{0.25mm} \delta_l \,. \hspace{4mm}
\end{split}
\end{eqnarray}
Notice that the fourth-order expansion is necessary to faithfully represent the dependence of the squared MEs on the Wilson coefficient set $C$ corresponding to the structure of the relevant Feynman diagrams shown in Figures~\ref{fig:diagramsSMEFT1} and~\ref{fig:diagramsSMEFT2}. Since the SM limit of the SMEFT expansion serves as the denominator hypothesis, the leading term in~(\ref{eq:taylor}) simplifies to unity, $a_0 (x) = 1$ for all $x$, making this choice both natural and practically advantageous. A set of NNs can then learn via regression the remaining non-trivial terms of the expansion, $a_{i,ij,ijk,ijkl}(x)$, on a per-event basis. Also observe that the ratio of rates $\sigma(\theta)/\sigma(\theta_0)$ can be expressed as an expectation value under the denominator hypothesis, 
\beq \label{eq:rateratio}
\frac{\sigma(\theta)}{\sigma(\theta_0)} = \int \! dx \; p \left (x|\theta_0 \right ) R \left (x; \theta - \theta_0 \right ) \,, 
\eeq 
which can subsequently be factored out of~(\ref{eq:taylor}) to derive the probability ratio. 

\begin{table*}[t!]
\centering
\renewcommand{\arraystretch}{1.5}
\begin{tabular}{w{c}{0.125\textwidth}|w{c}{0.125\textwidth}|w{c}{0.25\textwidth}|w{c}{0.25\textwidth}}
\toprule
\toprule
\multirow{2}{*}{Input features} & {$4\ell$} & \multicolumn{2}{c}{$\displaystyle \qty{ p_\ell = \left(p_T^{\ell}, \eta_{\ell}, \phi_{\ell}, E_{\ell}\right) }_{\ell=1,2,3,4} $}\\
{} & {$2 \ell 2\nu$} & \multicolumn{2}{c}{$\displaystyle \qty{ p_\ell = \left( E_{\ell}, p_T^{\ell}, \eta_{\ell}, \phi_{\ell} \right) }_{\ell=1,2} \,,\ \vec{E}_{T}^{\hspace{0.25mm}\mathrm{miss}} = \qty( E_{T}^{\hspace{0.25mm}\mathrm{miss}}, 
\phi_{\mathrm{miss}}) $}\\
\midrule
\multicolumn{2}{c|}{Numerator hypothesis} & {$gg\to (h^{\ast}\to)\,ZZ$} & {$q\bar{q} \to ZZ$, $q\bar{q} \to W^{+}W^{-}$} \\
\multicolumn{2}{c|}{Denominator hypothesis} & \multicolumn{2}{c}{SM $gg\to (h^{\ast}\to)\,ZZ$ process} \\
\multicolumn{2}{c|}{Parameter dependence} & {$C = \big\{C_H, C_{tH}, C_{HG}\big\}$} & {---}\\
\multicolumn{2}{c|}{Probability ratio} & { $R\left (x ; C \right )$ } & { $r \left (x ; q, g \right )$ } \\
\multicolumn{2}{c|}{Training target} & { $a_{i,ij,ijk,ijkl}(x)$ } & { $s(x)$ } \\
\multicolumn{2}{c|}{Output activation} & {linear} & {sigmoid}\\
\multicolumn{2}{c|}{Loss function} & {MSE} & {BCE}\\
\midrule
\multicolumn{2}{c|}{Number of layers times nodes} & \multicolumn{2}{c}{$20 \times 100$} \\
\multicolumn{2}{c|}{Batch size} & \multicolumn{2}{c}{$1024$}\\
\multicolumn{2}{c|}{Learning rate} & \multicolumn{2}{c}{$\leq 10^{-3}$}\\
\multicolumn{2}{c|}{Number of epochs} & \multicolumn{2}{c}{$\leq 300$}\\
\bottomrule
\bottomrule
\end{tabular}
\vspace{4mm}
\caption{NN structure and training details used for estimating likelihood ratios in our NSBI analysis. The symbol ``---'' indicates that the tree-level continuum background processes $q\bar q \to ZZ$ and $q\bar q \to W^+W^-$ are independent of the Wilson coefficients associated with the operators introduced in~(\ref{eq:operators}). The~coefficients $a_{i,ij,ijk,ijkl} (x)$ appear in the Taylor expansion~(\ref{eq:taylor}) of the rate-normalized likelihood~(\ref{eq:fratio}), while the learning rate is initially set to $10^{-3}$ and subsequently reduced depending on the training progress. Additional details are discussed in the main text.} 
\label{tab:nn_setup}
\end{table*}

Unlike in~(\ref{eq:prob_pp_mixture_model}), the NNs do not estimate probabilities, but rather their ratio relative to a chosen denominator hypothesis.
Since both the classifier and regression approaches use the SM $gg \to (h^{\ast} \to) \,ZZ$ process as the common denominator, however, the resulting ratios can still be consistently combined within the mixture model as 
\beq \label{eq:probability_model_pp_bsm_over_gg_sm}
\frac{p \left (x | C \right ) }{ g \left (x | 0 \right ) } = \frac{ \sigma_{g}(0) \hspace{0.25mm} R \left (x; C \right ) + \sigma_{q} \hspace{0.5mm} r(x) }{ \sigma_{g}(C) + \sigma_{q} } \,.
\eeq 

Finally, the inclusive probability ratio between two parameter points --- such as a BSM scenario versus the SM --- can be obtained by cancelling the common denominator
\begin{eqnarray} \label{eq:probability_model_pp_bsm_over_pp_sm}
\begin{split}
\frac{p \left (x | C \right ) }{ p \left (x | 0 \right ) } & = \frac{p\left (x | C \right ) / g \left (x | 0 \right )}{p \left (x|0 \right ) / g \left (x | 0 \right )} \\[2mm]
& \hspace{-1.25cm} = \frac{\sigma_{g}(0) + \sigma_{q}}{\sigma_{g}(C) + \sigma_{q}} \, \frac{ \sigma_{g}(0) \hspace{0.25mm} R \left (x; C \right ) + \sigma_{q} \hspace{0.5mm} r(x) }{ \sigma_{g}(0) + \sigma_{q} \hspace{0.5mm} r(x) } \,, \hspace{4mm}
\end{split}
\end{eqnarray} 
where $r(x)$ and $R \left (x ; C \right )$ are defined analogously to~(\ref{eq:carl}),~(\ref{eq:fratio}), and~(\ref{eq:taylor}), and approximated by their NN counterparts, $\hat{r} (x)$ and $\hat R (x; C)$, as explained in the next subsection. In turn, evaluating the product of these probability ratios over a set of events, $\qty{x_i}$, constitutes the shape term of the likelihood ratio in~(\ref{eq:likeratio}).

To the best of our knowledge, the multi-pronged approach described above and illustrated in~Figure~\ref{fig:mixture_model} is novel: it~combines the advantages of different existing NSBI methods while circumventing their challenges~\cite{cranmer2016approximatinglikelihoodratioscalibrated,Brehmer:2018hga,Brehmer:2018eca,Stoye:2018ovl}. Direct training on joint probability ratios, as performed in our study for the $gg \to (h^{\ast}\to) \, ZZ$ process over the BSM parameter space, is feasible only when~ME information can be readily extracted from a MC generator. For instance, calculating these quantities between the $q\bar{q}$- and $gg$-initiated processes is impractical because it involves convolutions with the PDFs. When~available, however, the joint probability ratios can dramatically improve the convergence and accuracy of the~NNs, as previously demonstrated in~\cite{Brehmer:2018hga,Brehmer:2018eca}. By factorising the contributions to the inclusive probability via the mixture model --- akin to the approach in~\cite{cranmer2016approximatinglikelihoodratioscalibrated,ATLAS:2024ynn} --- and estimating each ratio using the appropriate method at hand, we avoid these practical limitations associated with either of the NSBI techniques. In~addition, while {\tt MadMiner}~\cite{Brehmer:2019xox} integrates NSBI with {\tt MadGraph5\_aMC}~\cite{Alwall:2014hca}, our work extends this framework by enabling the extraction and utilization of joint probability ratios from {\tt MCFM~10.3} as well.

\subsection{NN training}
\label{sec:nn_training}

The set of observables $x$ used to train the NNs in this study is summarized in~\cref{tab:nn_setup}. For~leptons, the input features consist of their four-momentum components, expressed in terms of energy $E$, transverse momentum $p_T$, pseudorapidity $\eta$, and azimuthal angle $\phi$. For neutrinos, the event's missing transverse energy vector is used, characterized by its magnitude and azimuthal angle. To generate a large dataset $x_i$ of observables and the corresponding parton-level momenta $z_i$, a modified version of {\tt MCFM~10.3}, as described in~\cref{sec:MC}, is employed. The same MC generator also provides the squared MEs, $\abs{{\cal M} \left (z ; C \right )}^2$, for given parton-level momenta, $z$, and a choice of Wilson coefficients, $C$. This enables the calculation of the joint probability ratio for all events
\beq \label{eq:rMCFM}
\frac{p \left (x,z | C \right )}{p \left (x,z | 0 \right )} = \frac{\sigma(0)}{\sigma(C)} \frac{\abs{{\cal M} \left (z ; C \right )}^2}{\abs{{\cal M} \left (z ; 0 \right )}^2} \,,
\eeq
where the intractable latent-to-observable sampling processes cancel between hypotheses. Importantly, if the events are sampled according to 
\beq \label{eq:sampling}
x_i \sim p \left (x,z | 0 \right ) \,, 
\eeq
the regressed function, $\hat R(x)$, can be shown~\cite{Brehmer:2018eca,Stoye:2018ovl} to approximate the true joint probability ratio~(\ref{eq:rMCFM}) as a function of experimental observables, $x$.

An overview of the NN architectures and training configurations used in our NSBI analysis of the process $pp \to ZZ$ is provided in~\cref{tab:nn_setup}. The~implementation is written in {\tt pytorch}~\cite{Paszke:2019xhz} using the {\tt lightning}~\cite{jirka_borovec_2022_7447212} framework. The~classifier is trained on an equal number of samples drawn from the numerator and denominator hypotheses, labeled as $s(x) = 0$ and $s(x) = 1$, respectively. It~minimizes the binary cross-entropy~(BCE) loss, with a sigmoid activation function applied at the output node. Following~(\ref{eq:sampling}), the regressor is trained exclusively on events generated according to the SM $gg\to (h^{\ast}\to) \, ZZ$ process, using a linear output node in combination with a mean squared error~(MSE) loss. Hyperparameter optimization showed that increasing the network's depth and width improves performance, which saturates for architectures with 20 layers and 100 nodes per layer. The~hidden nodes employ the Swish activation function~\cite{DBLP:journals/corr/abs-1710-05941}. Following the application of analysis cuts (see~\cref{sec:event_selection}), approximately 3~million events are generated for each of the $gg \to (h^\ast \to) \, ZZ$ and $q\bar{q} \to ZZ$ processes contributing to the $4\ell$ final state. For the $2\ell 2\nu$ final state, the combined number of events from these processes and $q\bar{q} \to W^{+}W^{-}$ is lower, totalling around 1~million. The dataset is split into $75\%$ for training and $25\%$ for validation. Training is conducted with a batch size of 1024 at an initial learning rate of~$10^{-3}$, which is decreased by a factor of 10 if the validation loss does not improve over 5 consecutive epochs. If no further improvement is seen within~20 epochs, training is terminated early, with an upper limit of~300 epochs. 

\subsection{NN calibration}
\label{sec:nn_calibration}

\begin{figure}[!t]
\centering
\includegraphics[width=0.5\textwidth]{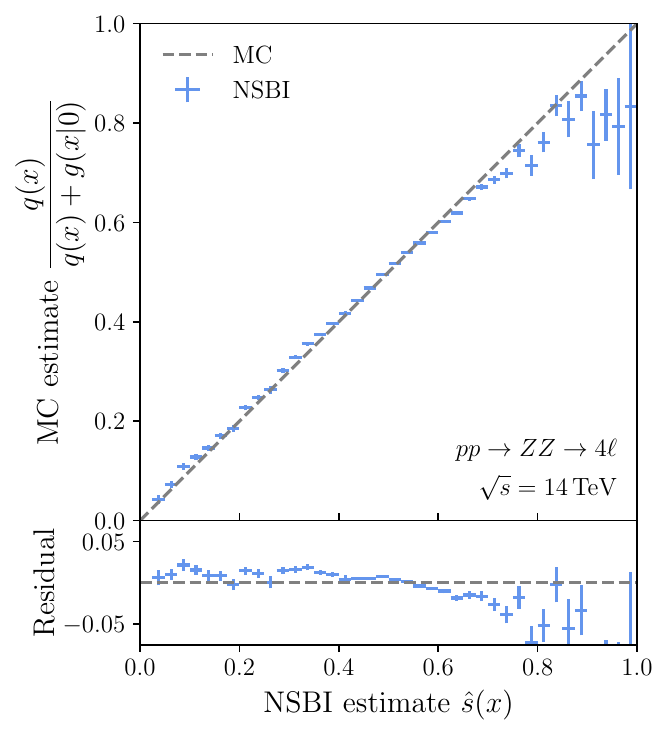}
\vspace{-2mm}
\caption{Calibration closure of the NN-estimated probability ratio. The residuals, shown below, represent the absolute difference between the predicted values $\hat s (x)$. Error bars reflect the statistical uncertainty arising from the finite size of the MC~sample.} \label{fig:carl_calibration}
\end{figure}

Reliable probability ratio estimates are critical for the robustness of our sensitivity study performed in~\cref{sec:results}. To assess the accuracy of the trained NNs in performing this task, we carry out a series of diagnostic tests, following approaches similar to those used by recent work by the ATLAS~collaboration~in~\cite{ATLAS:2024jry,ATLAS:2024ynn}. All evaluations are performed on datasets that are statistically independent from those used in training and validation.

For the probability ratio in~(\ref{eq:carl}), we compare the classifier decision function outputs, $\hat{s}(x)$, with the fraction of events originating from the numerator hypothesis among those sampled from the balanced hypothesis, as shown in~\cref{fig:carl_calibration}. For a well-calibrated classifier, the calibration curve appears as a diagonal line in the upper panel of the plot. The observed NSBI results closely follow this trend, indicating good calibration within statistical uncertainties. A corresponding calibration closure test for the $pp \to ZZ \to 2\ell 2\nu$ training is presented in~Appendix~\ref{sec:llnunuDiagnostics}.

\begin{figure}[!t]
\centering
\includegraphics[width=0.5\textwidth]{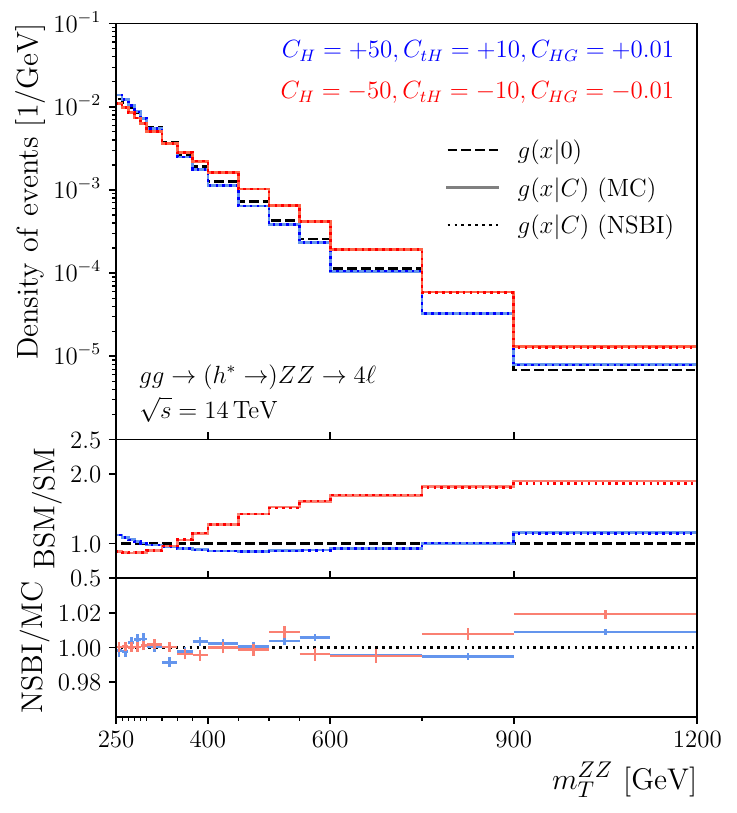}
\vspace{-2mm}
\caption{Reweighting closure comparing the NN-estimated Taylor expansion of BSM probability densities around the SM in the case of the $gg \to (h^\ast \to) \, ZZ \to 2 \ell 2 \nu$ process. The two BSM benchmarks are defined by $C_{H} = \pm 50$, $C_{tH} = \pm 10$, $C_{HG} = \pm 0.01$, and $\Lambda = 1 \, {\rm TeV}$. The error~bars in the lower panel represent the statistical uncertainty arising from the finite sample size of generated MC~events.} \label{fig:taylr_reweight}
\end{figure}

For the Taylor expansion in~(\ref{eq:taylor}), \cref{fig:taylr_reweight} presents a comparison of reweighting from the SM event distribution to that of two BSM scenarios --- defined by $C_{H} = \pm 50$, $C_{tH} = \pm 10$, $C_{HG} = \pm 0.01$, and $\Lambda = 1 \, {\rm TeV}$ --- in the $gg \to (h^\ast \to) \, ZZ \to 2 \ell 2 \nu$ channel using two methods: first, the parton-level squared ME ratio, and, second, the density ratio estimated by the~NNs. This comparison is carried out differentially in the transverse mass $m_{T}^{ZZ}$, as defined in~(\ref{eq:mTZZ}), and shows excellent agreement, indicating strong closure of the NSBI estimate.

Finally, the expectation value of the inclusive $pp \to ZZ$ probability ratio, defined under the mixture model in~(\ref{eq:probability_model_pp_bsm_over_pp_sm}), over the SM denominator hypothesis is computed: 
\begin{eqnarray} \label{eq:inclusivecalibration}
\left \langle \frac{ p \left (x | C \right ) }{ p \left (x | 0 \right )}\right \rangle_{{\rm SM}} = \int \! d{ x} \; \frac{ p \left (x | C \right ) }{ p \left (x | 0 \right )} \; p \left (x | 0 \right ) \,. \hspace{5mm} 
\end{eqnarray} 
With perfect training and infinite statistics, the expectation value (\ref{eq:inclusivecalibration}) equals~1 for any Wilson coefficient set $C$ tested. These deviations are corrected for by rescaling the probability ratio as follows
\beq \label{eq:lr_calibration}
\frac{ p \left (x | C \right ) }{ p \left (x | 0 \right )} \, \mapsto \, \left \langle \frac{ p \left (x | C \right ) }{ p \left (x | 0 \right )}\right \rangle_{\mathrm{SM}}^{-1} \hspace{1mm} \frac{ p \left (x | C \right ) }{ p \left (x | 0 \right )} \,,
\eeq
that is, we multiply by the inverse of the expectation value defined in~(\ref{eq:inclusivecalibration}). For a given dataset with finite sample size, this procedure guarantees that the best-fit parameter value in the NSBI framework converges to the true value, i.e., $\hat C = C$, as will be demonstrated explicitly in~\cref{fig:ci_nsbi_vs_rate_c6}. In other words, all results presented below in~\cref{sec:offshell_ppZZ_sensitivity} correspond to Asimov datasets, where the test datasets are generated exactly according to the assumed hypotheses (SM~or an alternative) at a finite sample size. When~this approach is applied in an experiment, the NSBI estimates should be made robust with ensembling techniques~\cite{Brandes:2024vhw,ATLAS:2024ynn}, the uncertainty due to the finite sample size should be incorporated~\cite{Benevedes:2025nzr} and the test statistic may be calibrated using a Neyman construction~\cite{ATLAS:2024ynn}.

\section{Sensitivity study} 
\label{sec:results}

In this section, we conduct a comprehensive sensitivity analysis of off-shell Higgs production via the ggF process at the HL-LHC. Building upon the NSBI framework introduced in the previous section, we place constraints on the Wilson coefficient $C_H$ individually, as well as on all pairwise combinations of the Wilson coefficients $C_H$, $C_{tH}$, and $C_{HG}$. To fully exploit the sensitivity of the $pp \to ZZ$ channel, our analysis includes both the~$4\ell$ and $2\ell 2\nu$ final states.

As already noted above, we do not account for uncertainties arising from finite sample sizes or their potential impact on the NSBI model misspecification. In the following, we also exclude both experimental and theoretical systematic uncertainties. Experimentally, the dominant systematic uncertainties come from the jet energy scale and resolution. On the theoretical side, systematic uncertainties stem from the choice of~PDFs, missing higher-order corrections in both QCD and EW perturbative calculations, and the modeling of the parton shower. In the ATLAS measurement of the off-shell cross section during LHC~Run~2~\cite{ATLAS:2024jry}, combined systematic uncertainties are roughly~$3.5$ times smaller than the statistical uncertainties. Extrapolating these results to the full HL-LHC luminosity indicates that systematic uncertainties could become comparable to the statistical sensitivity presented below, assuming no further improvements in experimental or theoretical precision. As a result, in a realistic HL-LHC analysis, systematic uncertainties would need to be incorporated comprehensively. However, this is beyond the scope of the present work.

\subsection{Event selection}
\label{sec:event_selection}

\begin{table*}[t!]
\centering
\begin{tabular}{c | c | c | c | c}
\toprule
\toprule
\multirow{2}{*}{Event selection} & \multicolumn{4}{c}{Cross section $\mathrm{[\si{\femto\barn}]}$}\\
{} & {{$gg \to h^{\ast} \to ZZ$}} & {$gg \to (h^\ast \to) \; ZZ$} & {$q \bar{q} \to ZZ$} & {$q\bar{q} \to W^+ W^-$}\\
\midrule
\multicolumn{5}{c}{$pp \to 4\ell$}\\
\midrule
{$p_{T}^{\ell_{1,2,3,4}} > 20, 15, 10, 7 \, {\rm GeV}$} & \multirow{3}{*}{$0.3099(4)$} & \multirow{3}{*}{$5.656(5)$} & \multirow{3}{*}{44.87(3)} & \multirow{3}{*}{---}\\
{$70 \, {\rm GeV} < m_{\ell\ell} < 110 \, {\rm GeV}$} & {} & {} & {}\\
{$m_{ZZ} > 180 \, {\rm GeV}$} & {} & {} & {}\\
\midrule
\multicolumn{5}{c}{$\HepProcess{pp \to 2\ell 2\nu}$}\\
\midrule
{$p_{T}^{\ell_{1,2}} > 30, 20 \, {\rm GeV}$} & \multirow{5}{*}{1.4006(9)} & \multirow{5}{*}{5.046(7)} & \multirow{5}{*}{44.49(6)} & \multirow{5}{*}{1.47(1)}\\
{$80 \, {\rm GeV} < m_{\ell\ell} < 100 \, {\rm GeV}$} & {} & {} & {} & {}\\
{$\Delta R_{\ell \ell} < 2$} & {} & {} & {} & {}\\
{$E_{T}^{\mathrm{miss}} > 60 \, {\rm GeV}$} & {} & {} & {} & {}\\
{$m_{T}^{ZZ} > 250 \, {\rm GeV}$} & {} & {} & {} & {}\\
\bottomrule
\bottomrule
\end{tabular}
\vspace{4mm}
\caption{Summary of event selection criteria applied in the $4 \ell$ and $2\ell 2 \nu$ final-state analyses. All selected leptons are required to have pseudorapidity $|\eta_\ell| < 2.5$. The table also presents the fiducial cross section contributions from the considered signal and background processes within the SM, assuming $pp$ collisions at $\sqrt{s} = 14 \, {\rm TeV}$. The number in parentheses indicate the statistical uncertainty at the last reported decimal place, and the symbol ``---'' indicates that, after applying the selection cuts, the contribution from $q \bar q \to W^+ W^-$ to the $4 \ell$ final state is negligibly small.}
\label{tab:event_preselection}
\end{table*}

The events considered in our analysis consist of leptons and neutrinos originating from the decays of the $Z$ bosons. For the $pp \to ZZ \to 4 \ell$ process, we apply selection criteria on $m_{ZZ}$, as well as on the pseudorapidity $\eta$ and transverse momentum $p_T$ of the leptons, as previously described in~\cref{sec:distributions}. In the case of same-flavor final states ($2 e^+ 2 e^-$ and $2 \mu^+ 2 \mu^-$), the two $Z$-boson candidates are reconstructed by pairing opposite-sign leptons such that the resulting invariant mass~$m_{\ell \ell}$ is closest to the nominal $Z$-boson mass. Additionally, we require the invariant mass to lie within the window $70 \, {\rm GeV} < m_{\ell \ell} < 110 \, {\rm GeV}$. 

For the $pp \to ZZ \to 2 \ell 2 \nu$ channel, the selected leptons are required to satisfy $|\eta_\ell| < 2.5$, with transverse momentum thresholds of $p_{T}^{\ell_1} > 30 \, {\rm GeV}$ for the leading lepton and $p_{T}^{\ell_2} > 20 \, {\rm GeV}$ for the subleading one. The invariant mass of the dilepton pair must lie within $80 \, {\rm GeV} < m_{\ell \ell} < 100 \, {\rm GeV}$ and their angular separation must fulfill $\Delta R_{\ell \ell} < 2$. Additionally, events must exhibit missing transverse energy of $E_{T}^{\mathrm{miss}} > 60 \, {\rm GeV}$, and the transverse mass of the system must satisfy $m_{T}^{ZZ} > 250 \, {\rm GeV}$. The transverse mass is defined~as 
\begin{align} \label{eq:mTZZ}
m_{T}^{ZZ} & = \Bigg ( \left [ \sqrt{m_Z^2 + \left (p_{T}^{\ell \ell} \right )^2} + \sqrt{m_Z^2 + \left (E_{T}^{\mathrm{miss}} \right )^2} \right ]^2 \nonumber \\
& \phantom{xxx} - \left | \vec{p}_{T}^{\hspace{0.75mm} \ell \ell} + \vec{E}_{T}^{\hspace{0.25mm} \mathrm{miss}} \right |^2 \Bigg )^{1/2} \,,
\end{align}
where $m_Z$ denotes the mass of the $Z$ boson, $p_{T}^{\ell\ell}$ is the transverse momentum of the dilepton system, with $\vec{p}_{T}^{\hspace{0.75mm} \ell \ell}$ representing its vector form, and $\vec{E}_{T}^{\hspace{0.25mm} \mathrm{miss}}$ corresponds to the transverse momentum vector of the dineutrino system.

Table~\ref{tab:event_preselection} summarizes the selection criteria applied to the two final states resulting from $pp \to ZZ$, along with the corresponding fiducial cross section contributions from the signal and background processes in the SM. The given values correspond to two lepton flavors, $\ell = e, \mu$, and include the $K$-factors specified in~\cref{sec:distributions}, corresponding to $\sqrt{s} = 14 \, {\rm TeV}$. For the signal process $gg \to h^\ast \to ZZ$, the contribution to the $pp \to 2 \ell 2 \nu$ channel is about $4.5$ times larger than that to $pp \to 4 \ell$, primarily due to the $Z \to \nu \bar \nu$ branching ratio being roughly three times greater than $Z \to \ell^+ \ell^-$, with ${\rm BR} (Z \to \nu \bar \nu) = 20.00\%$ and ${\rm BR} (Z \to \ell^+ \ell^-) = 6.73\%$~\cite{ParticleDataGroup:2024cfk}. However, once the continuum box background and interference contributions are included in the ggF process, this enhancement is significantly reduced, resulting in similar fiducial rates for $gg \to (h^\ast \to) \, ZZ$ in both final states. Nevertheless, the table clearly shows that including both the $4 \ell$ and $2\ell 2\nu$ channels doubles the event rate compared to considering $pp \to 4 \ell$ alone, thereby fully leveraging the sensitivity of the $pp \to ZZ$ channel. It is also noteworthy that the non-interfering background from $q \bar q \to ZZ$ exceeds the corresponding $gg \to (h^\ast \to) \, ZZ$ contribution by more than a factor of~$8$ in both final states. The background from $q \bar q \to W^+ W^-$ is negligible for the $4 \ell$ final state and subleading for the $2 \ell 2 \nu$ channel under the event selection criteria used in our analysis.

\subsection{Constraints based on NSBI analysis of $gg \to (h^\ast \to) \, ZZ$} 
\label{sec:offshell_ggZZ_sensitivity}

\begin{figure}[t!]
\centering
\includegraphics[width=0.49\textwidth]{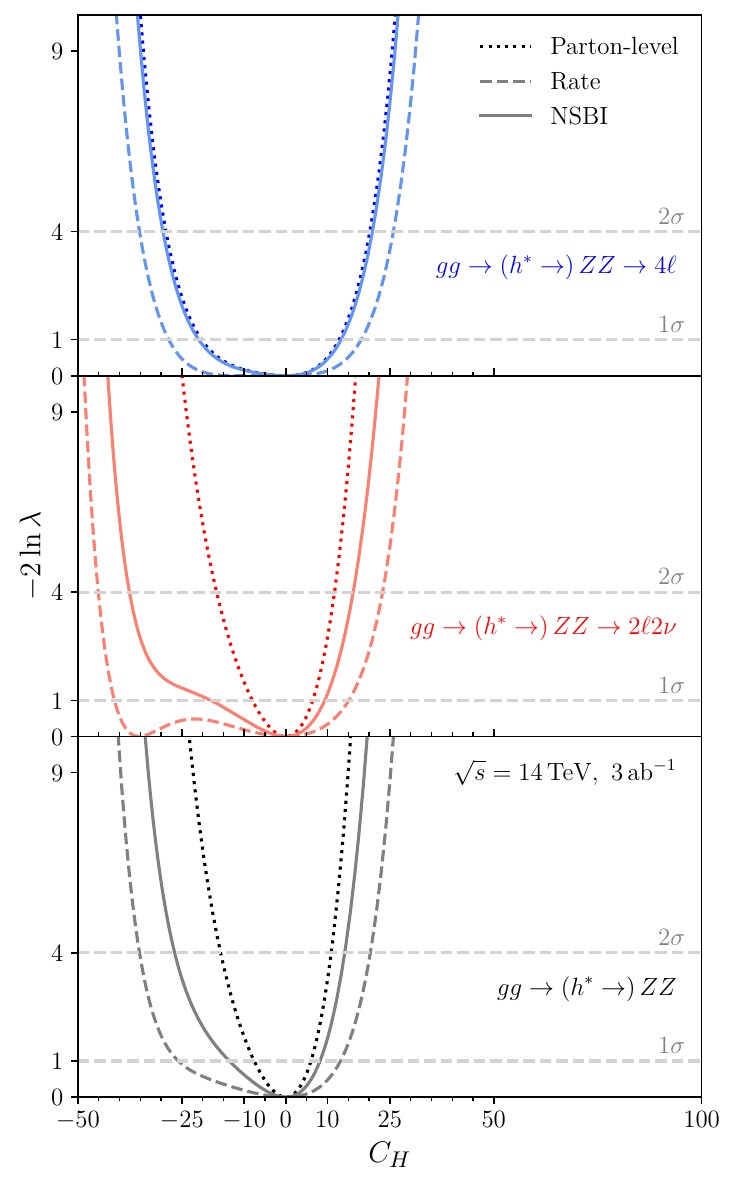}
\vspace{-2mm}
\caption{The negative log-likelihood $-2 \ln \lambda$, plotted as a function of the Wilson coefficient~$C_H$. The upper, middle, and lower panels present the results for the $4\ell$ and $2 \ell 2 \nu$ final states, as well as their combination, respectively, in the $gg \to (h^\ast \to ) \, ZZ$ channel. The lower panel displays the combined result from both final states. In each panel, constraints from a parton-level analysis~(dotted lines), a rate-only measurement~(dashed lines), and the NSBI analysis~(solid lines) are shown.} \label{fig:negll_4l_2l2v}
\end{figure}

\cref{fig:negll_4l_2l2v} presents the negative log-likelihood as a function of the Wilson coefficient $C_H$, as derived from our NSBI analysis. The upper and middle panels show results for the $gg \to (h^\ast \to) \, ZZ \to 4\ell$ and $gg \to (h^\ast \to) \, ZZ \to 2\ell 2\nu$ channels, respectively. The lower panel combines the individual negative log-likelihoods to yield the constraints for the full $gg \to (h^\ast \to) \, ZZ$ process. The analysis is carried out under HL-LHC conditions, assuming $pp$ collisions at $\sqrt{s} = 14 \, {\rm TeV}$ and an integrated luminosity of $3 \, {\rm ab}^{-1}$. For comparison, we also include negative log-likelihoods from a parton-level analysis --- based on the joint probability ratios in~(\ref{eq:rMCFM}) --- as well as from a rate-only analysis that considers only the Poisson term in~(\ref{eq:likelihood}). In~the case of the $4\ell$ final state (upper panel), it is evident that the NSBI approach --- by exploiting the full event kinematics as a proxy of the MEs --- closely approaches the optimal parton-level performance. This is expected given that in the $4\ell$~channel, all final-state particles are precisely measurable and the Higgs rest frame can be fully reconstructed. Therefore, the inclusion of shape information through NSBI significantly enhances the sensitivity to $C_H$ compared to using rate-only~data.

A different picture emerges for the $2\ell 2\nu$ final state, shown in the middle panel. Given the larger fraction of Higgs signal relative to the continuum backgrounds, one can expect even better constraints at the parton level. The NSBI analysis, however, cannot not achieve this sensitivity. This limitation arises because in the $2\ell 2\nu$ channel, only one $Z$-boson candidate can be reconstructed from the dilepton system, while the neutrinos contribute solely in the form of $\vec{E}_{T}^{\hspace{0.25mm}\mathrm{miss}}$. This leads to an incomplete representation of the full event kinematics, effectively projecting the high-dimensional latent space onto a lower-dimensional observable space. The figure clearly illustrates that this inevitable loss of information has a significant impact in this case, resulting in the NSBI-derived negative log-likelihood curve that lies between those obtained from the parton-level and rate-only analyses. Nevertheless, the $2\ell 2\nu$ channel provides a notable improvement over the $4\ell$ channel in excluding $C_H > 0$ values, while for $C_H <0$, the constraint is comparable up to around the $1\sigma$~confidence level~(CL).

The lower panel of~\cref{fig:negll_4l_2l2v} shows the negative log-likelihood resulting from the combination of both final states. As before, the apparent lower sensitivity of the NSBI analysis relative to the parton-level study stems from the incomplete experimental reconstruction of the full final-state kinematics in the $gg \to (h^\ast \to) \, ZZ \to 2\ell 2\nu$ channel. Our NSBI analysis of the process $gg \to (h^\ast \to) \, ZZ$ leads to the bound $-26.7 < C_H < 13.9$ on the Wilson coefficient of the operator $Q_H$ at~$2 \sigma$~CL. Applying~(\ref{eq:kappaparameter}), this limit corresponds to a~$2 \sigma$~CL constraint $-5.5 < \kappa_\lambda < 13.5$. Leveraging the full LHC~Run~2 dataset at $\sqrt{s} = 13 \, {\rm TeV}$ of $140 \, {\rm fb}^{-1}$ of integrated luminosity, both ATLAS and CMS have set limits on~$\kappa_\lambda$ using inclusive analyses of a wide array of on-shell processes involving double- and single-Higgs production~\cite{ATLAS:2022jtk,CMS:2024awa}. Under the assumption that all other Higgs couplings retain their SM values, the derived $2 \sigma$~CL bounds on the coupling modifier $\kappa_\lambda$ are $-1.4 < \kappa_\lambda < 6.1$ and $-1.2 < \kappa_\lambda < 7.5$, respectively. This demonstrates that off-shell Higgs production in the ggF channel offers inferior sensitivity to $\kappa_\lambda$ or equivalently~$C_H$ relative to double-Higgs production in a single-parameter analysis that neglects the non-interfering continuum background from $q \bar{q} \to ZZ$. The strength of the $pp \to ZZ$ channel, as we will explain in the next subsection, arises from the different way this process depends on the three SMEFT operators introduced in~(\ref{eq:operators}) relative to $pp \to hh$ production. This feature is essential for breaking degeneracies when simultaneously analyzing multiple Wilson coefficients in a combined Higgs physics analysis.

\subsection{Constraints based on NSBI analysis of $pp \to ZZ$} 
\label{sec:offshell_ppZZ_sensitivity}

\begin{figure}[t!]
\centering
\includegraphics[width=0.49\textwidth]{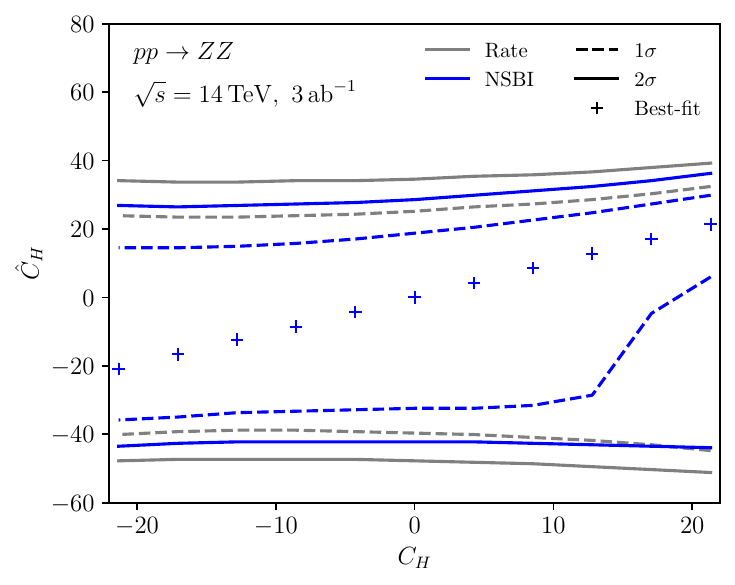}
\vspace{-2mm}
\caption{The best-fit values $\hat C_H$ along with the $\pm 1\sigma$ and $\pm 2\sigma$~CL intervals as a function of the true $C_H$, using Asimov datasets generated in the range $-20$ to $+20$. Results from the NSBI and rate-only analyses are represented by the blue and gray curves, respectively.} \label{fig:ci_nsbi_vs_rate_c6}
\end{figure}

\begin{figure*}[t!]
\centering
\includegraphics[width=0.475\textwidth]{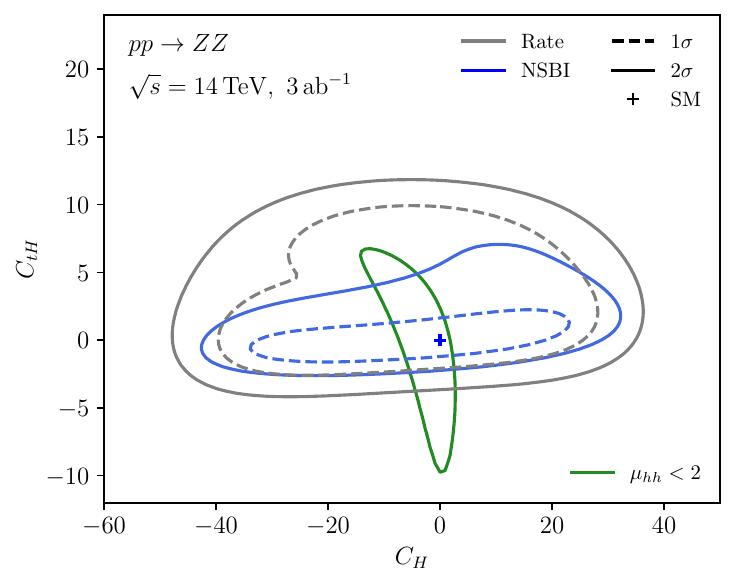} \quad 
\includegraphics[width=0.475\textwidth]{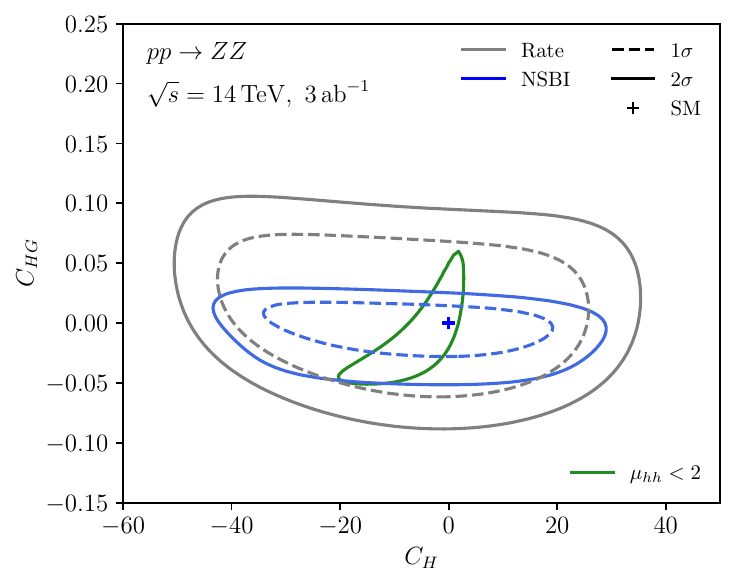} \\[2mm]
\includegraphics[width=0.475\textwidth]{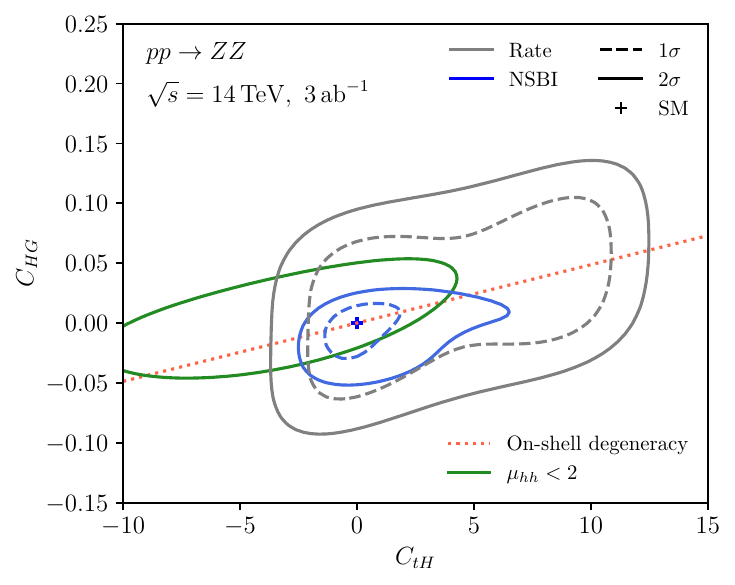}
\vspace{4mm}
\caption{ Constraints in the $C_H \hspace{0.25mm}$--$\hspace{0.5mm} C_{tH}$, $C_H \hspace{0.25mm}$--$\hspace{0.5mm} C_{HG}$, and $C_{tH} \hspace{0.25mm}$--$\hspace{0.5mm} C_{HG}$ parameter planes derived from our NSBI analysis of the $pp \to ZZ$ process are shown by the blue contours. These results are based on $pp$ collisions at $\sqrt{s} = 14 \, {\rm TeV}$ with an integrated luminosity of $3 \, {\rm ab}^{-1}$. For comparison, constraints from rate-only measurements are indicated by the gray contours. In all cases, the third Wilson coefficient not displayed is set to zero, and the results are obtained assuming $\Lambda = 1 \, {\rm TeV}$. The solid green contours in all panels show the constraints obtained from a signal strength measurement of the double-Higgs production cross section with $\mu_{hh} < 2$. The dotted red line in the lower panel marks the flat direction, $C_{HG} = 4.9 \cdot 10^{-3} \, C_{tH}$, which cannot be constrained by on-shell Higgs measurements. Further details are provided in the text.} \label{fig:nll_pp}
\end{figure*}

\cref{fig:ci_nsbi_vs_rate_c6} presents the best-fit values $\hat C_H$ along with the $\pm 1\sigma$ and $\pm 2\sigma$~CL intervals as a function of the true $C_H$, using Asimov datasets generated in the range $-20$ to $+20$. The displayed results are obtained from the NSBI analysis of the full $pp \to ZZ$ process. In~\cref{fig:nll_pp}, the constraints derived from the NSBI analysis of the $pp \to ZZ$ process are shown in the three parameter planes formed by pairs of the Wilson coefficients $C_H$, $C_{tH}$, and $C_{HG}$, as defined in~(\ref{eq:operators}). The results shown in both figures are based on $pp$ collisions at $\sqrt{s} = 14 \, {\rm TeV}$ with an integrated luminosity of $3 \, {\rm ab}^{-1}$. To evaluate the enhancement offered by the NSBI analysis relative to rate-only measurements, the corresponding rate-only constraints are also included for comparison. In each panel, the Wilson coefficient not shown is set to zero, and all dimension-six SMEFT contributions assume a common suppression scale of $\Lambda = 1 \, {\rm TeV}$.

Compared to~\cref{fig:negll_4l_2l2v},~\cref{fig:ci_nsbi_vs_rate_c6} shows a notable reduction in constraining power for an assumed Asimov value of $C_H = 0$, primarily due to the substantial contamination from $C_H$-independent $q\bar{q} \to ZZ$ events in the signal region of the analysis. Consequently, the NSBI analysis provides only a modest improvement over the rate-only measurement in constraining the Wilson coefficient, $C_H$, or equivalent, $\kappa_\lambda$. At the HL-LHC, we project a $2 \sigma$~CL constraint of $-43.5 < C_H < 26.9$, which translates to $-11.6 < \kappa_\lambda < 21.4$. The effect of the tree-level continuum background $q \bar{q} \to ZZ$ can be mitigated by using more advanced multivariate techniques to improve the separation between $gg$- and $q\bar{q}$-initiated events to surpass the simple event selection criteria described in Table~\ref{tab:event_preselection}. Such methods have been applied~\cite{ATLAS:2023syu}, but we leave a detailed exploration of these approaches to a future analysis based on actual LHC data. Interestingly, the NSBI method shows a significant improvement in constraining power at the $1 \sigma$~CL for $C_H \gtrsim 13$, driven by the substantial changes in event kinematics associated with these scenarios, as illustrated earlier in~Figures~\ref{fig:m4lCH} and~\ref{fig:pTl1CH}. However, at the $2 \sigma$~CL, the sensitivity is still primarily governed by rate information. Notice that in every case, the best-fit value $\hat C_H$ aligns with the injected $C_H$ value in the Asimov dataset, providing a validation of our fitting procedure.

The~upper left panel of~\cref{fig:nll_pp} displays the constraints in the $C_H \hspace{0.25mm}$--$\hspace{0.5mm} C_{tH}$ plane. This behavior can be understood from the lower panels of~Figures~\ref{fig:m4lCH} and~\ref{fig:pTl1CH}, which show that significant changes in the distributions occur where the bulk of the cross section lies, while modifications in the tails are relatively small and featureless. In~the case of $C_{tH}$, incorporating the shape of the $pp \to ZZ$ distributions in the NSBI analysis yields a clear enhancement of the resulting constraints. It~is~worth noting that the improvement is more significant for positive values of $C_{tH}$ than for negative ones. Figures~\ref{fig:m4lCtHCHG} and~\ref{fig:pTl1CtHCHG} illustrate that this effect arises because positive (negative) $C_{tH}$ values suppress (enhance) the differential rate. Shape information is thus more relevant for $C_{tH} > 0$ than for $C_{tH} < 0$. A similar pattern is observed in the upper right panel, which shows the constraints in the $C_H \hspace{0.25mm}$--$\hspace{0.5mm} C_{HG}$ plane. However, contrary to the prior example, the shape information proves more crucial for positive than for negative values of the Wilson coefficient $C_{HG}$. This occurs because, for $C_{HG} > 0$, the alterations in the tails of the kinematic distributions are more pronounced than for $C_{HG} < 0$, as can be inferred from Figures~\ref{fig:m4lCtHCHG} and~\ref{fig:pTl1CtHCHG}.

Finally, the lower panel in~\cref{fig:nll_pp} shows the constraints in the $C_{tH} \hspace{0.25mm}$--$\hspace{0.5mm} C_{HG}$ plane. In this case including shape information through a NSBI analysis leads to a significant tightening of the constraints on both $C_{tH}$ and $C_{HG}$. In particular, the NSBI analysis of the off-shell $pp \to ZZ$ process can partially lift the flat direction $C_{HG} = 4.9 \cdot 10^{-3} \, C_{tH}$ --- indicated by the dotted red line in the lower panel of~\cref{fig:nll_pp} --- which remains inaccessible to on-shell Higgs measurements. That $pp \to h j$ and $pp \to ZZ$ production offers a mean to partially decouple the contributions from~$C_{tH}$ and~$C_{HG}$ was originally pointed out in~\cite{Azatov:2013xha,Grojean:2013nya} and~\cite{Azatov:2014jga,Buschmann:2014sia}, respectively. Constraints on the Wilson coefficients~$C_{tH}$ and~$C_{HG}$ using off-shell Higgs production in the ggF channel have already been derived by~ATLAS in~\cite{ATLAS:2023syu}.

To highlight the complementarity and interplay between Wilson coefficient determinations from off-shell single-Higgs and on-shell double-Higgs production, all panels of~\cref{fig:nll_pp} also display the constraints from a hypothetical signal strength measurement of the double-Higgs production cross section with $\mu_{hh} < 2$, shown as solid green contours. The quoted bound corresponds to the $95\,\%$~CL upper limit reported in the HL-LHC projection by~ATLAS~\cite{ATLAS:2022faz}, assuming the systematic uncertainties of LHC~Run~2~conditions. Our calculation of $\mu_{hh}$ employs the results given in \cite{Buchalla:2018yce}, which incorporate the full top-quark mass dependence for the $pp \to hh$ cross section at NLO in QCD, both within the SM and in the presence of the Wilson coefficients $C_H$, $C_{tH}$, and~$C_{HG}$. As~evident from all three panels, $pp \to hh$ offers substantially stronger constraints on $C_H$, whereas in the two-dimensional parameter space, the Wilson coefficients $C_{tH}$ and $C_{HG}$ are better constrained by our NSBI analysis of the $pp \to ZZ$ process. Note that although the double-Higgs constraint in the $C_{tH} \hspace{0.25mm}$--$\hspace{0.5mm} C_{HG}$ plane is relatively weak, it still helps to resolve the degeneracy present in on-shell Higgs production. This occurs because the $gg \to hh$ amplitude includes both linear and quadratic terms in $C_{tH}$, whereas it depends only linearly on $C_{HG}$. In particular, the quadratic term in $C_{tH}$ breaks the degeneracy, a feature that is absent in the $gg \to h$ amplitude. The discussion above implies that in a global SMEFT analysis involving the Higgs self-coupling, off-shell Higgs production through the ggF channel provides valuable information by helping to resolve flat directions in the multi-dimensional space of Wilson coefficients that on-shell production cross-section measurements alone cannot~constrain.

\FloatBarrier

\section{Conclusions}
\label{sec:conclusions}

This article explored the constraints on the Higgs trilinear self-coupling obtained from off-shell Higgs production at the HL-LHC, along with their relationship to constraints on other relevant SMEFT dimension-six operators. To tackle this, we have developed an NSBI framework for statistical inference that leverages NNs to estimate likelihood ratios, essentially implementing a machine learning-based version of the ME method used in~\cite{Haisch:2021hvy} for a similar analysis. 

The NNs were trained using squared MEs that model the Higgs signal --- including the dimension-six SMEFT operators introduced in~(\ref{eq:operators}) --- along with relevant background processes and quantum interference effects in the $pp \to ZZ \to 4 \ell$ and $pp \to ZZ \to 2 \ell 2 \nu$ channels. These components are derived using a modified version of the MC generator {\tt MCFM~10.3}~\cite{Boughezal:2016wmq}. The trained and calibrated NNs were then employed to estimate the inclusive probability ratio for a mixture model designed to fully capture the experimentally accessible kinematic information of the $pp \to ZZ$ process. Given the importance of reliable NN predictions for the robustness of our HL-LHC sensitivity study, we performed a series of diagnostic tests, inspired by similar validation procedures recently employed by the ATLAS collaboration in~\cite{ATLAS:2024jry,ATLAS:2024ynn}. The trained NNs demonstrated good performance in both classification and regression~tasks. We developed a method to consistently combine the two approaches for estimating the probability ratio of the inclusive $pp \to ZZ$ process between SM and BSM hypotheses, even with finite training data.

We then performed a detailed sensitivity analysis of off-shell Higgs production via ggF at the HL-LHC. To assess the effectiveness of the NSBI approach and evaluate its performance, we first focused on the $gg$-initiated channels only. In particular, we have shown that for $gg \to (h^\ast \to) \, ZZ \to 4 \ell$, our NSBI approach achieves near-optimal performance, closely matching the parton-level results. This outcome is expected because, in the $4 \ell$ channel, all final-state particles are precisely measurable, allowing for complete reconstruction of the Higgs rest frame. A different situation occurs for the $2 \ell 2 \nu$ final state, as the four-momenta of the neutrinos are not directly accessible at the LHC. Consequently, the NSBI analysis of $gg \to (h^\ast \to) \, ZZ \to 2 \ell 2 \nu$ is significantly less sensitive to SMEFT effects than an analysis performed at the parton level. In both cases, however, our NSBI approach outperforms analyses relying solely on rate data, highlighting the importance of incorporating shape information. 

Our attention then turned to the full $pp \to ZZ$ production channel. As a first step, we derived the HL-LHC projected single-parameter bound on the Wilson coefficient $C_H$, which encodes the leading SMEFT correction to the trilinear Higgs self-coupling modifier $\kappa_\lambda$. Assuming $\Lambda = 1 \, {\rm TeV}$, our NSBI analysis yields a $2 \sigma$~CL interval of $-43.5 < C_H < 26.9$. This constraint corresponds to $-11.6 < \kappa_\lambda < 21.4$, which is considerably weaker than the current bounds on $\kappa_\lambda$ reported by the ATLAS and CMS collaborations in~\cite{ATLAS:2022jtk,CMS:2024awa}. Undeterred by this finding, we proceeded to analyze the constraints in the three parameter planes formed by pairs of the Wilson coefficients $C_H$, $C_{tH}$, and $C_{HG}$. Our NSBI analysis shows that at the HL-LHC, off-shell Higgs production via $pp \to ZZ$ is expected to probe values of the Wilson coefficients $C_{tH}$ and $C_{HG}$ across all three two-dimensional parameter planes, accessing regions beyond the reach of cross-section measurements from on-shell single- and double-Higgs production in ggF. In a global SMEFT analysis, off-shell Higgs production can help resolve degeneracies in the Wilson coefficient space and should thus be included.

It is important to recall that the sensitivity of our NSBI analysis for the full $pp \to ZZ$ process is notably limited by the significant $q \bar q \to ZZ$ contamination in the signal region. This limitation could be alleviated by applying advanced multivariate techniques~\cite{ATLAS:2023syu} to improve discrimination between $gg$- and $q \bar q$-initiated events, beyond the basic selections employed in our work. Including $pp \to ZZ$ production with varying jet multiplicities and/or EW Higgs production channels is also expected to enhance the strength of the constraints.

Applying the NSBI approach to the processes $pp \to ZZ \to 4 \ell$ and $pp \to ZZ \to 2 \ell 2 \nu$, as demonstrated here, offers a relatively straightforward path to constraining Higgs-portal models~\cite{Haisch:2022rkm,Haisch:2023aiz} or placing bounds on the spectral density of the Higgs. Another promising phenomenological application at the LHC involves incorporating differential information from on-shell single-Higgs production processes, such as Higgs plus jet production~\cite{Haisch:2024nzv}, to provide complementary indirect constraints on the Higgs trilinear self-coupling. Ultimately, the latter efforts are directed toward enabling a comprehensive NSBI analysis of the Higgs potential, incorporating contributions from on-shell double-Higgs production~\cite{Mastandrea:2024irf}. We defer exploration of these directions to future~work.

\backmatter

\bmhead{Acknowledgements}

This project was initiated during the PHYSTAT-SBI 2024 - Simulation Based Inference in Fundamental Physics workshop, held from May 15-17, 2024, at the Max Planck Institute for Physics in Garching~\cite{PHYSTATSBI2024}. AG was supported by the DOE Office of Science. UH thanks Jeffrey~Davies, Andrei~Gritsan, Lazar~Markovic, Giacomo~Ortona, and Toni~Sculac for their interest in~\cite{Haisch:2021hvy} and for the valuable email correspondence regarding the $pp \to ZZ$ implementation in the {\tt JHUGen~MELA} package~\cite{Gritsan:2020pib}. PTH thanks John M. Campbell for his availability and assistance with the modifications to the~{\tt MCFM~10.3}~\cite{Boughezal:2016wmq} source code. The authors thank Gilles Louppe for his valuable feedback on the initial version of this manuscript. The Feynman diagrams shown in this article were drawn with the {\tt TikZ-Feynman} package~\cite{Ellis:2016jkw}. 

\appendix

\section{Additional kinematic distributions}
\label{sec:additionalc6Benchmarks}

\begin{figure}[t!]
\centering
\includegraphics[width=0.45\textwidth]{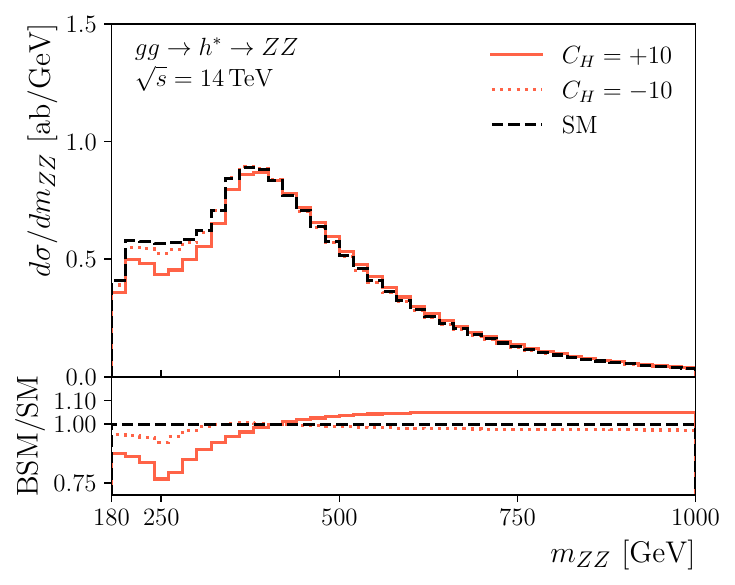}
\includegraphics[width=0.45\textwidth]{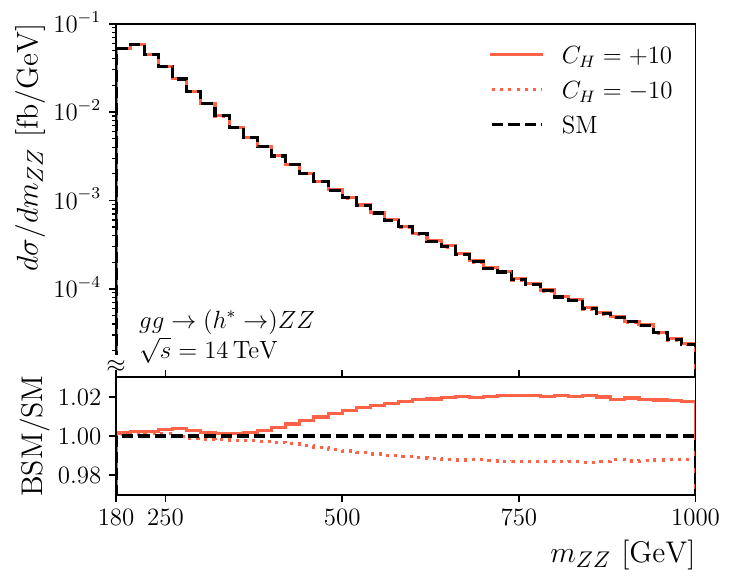}
\vspace{2mm}
\caption{\label{fig:m4lCH_10} Same as~\cref{fig:m4lCH}, but showing results for the two values $C_H = \pm 10$ of the Wilson coefficient appearing in~(\ref{eq:kappaparameter}).}
\end{figure}

\begin{figure}[t!]
\centering
\includegraphics[width=0.49\textwidth]{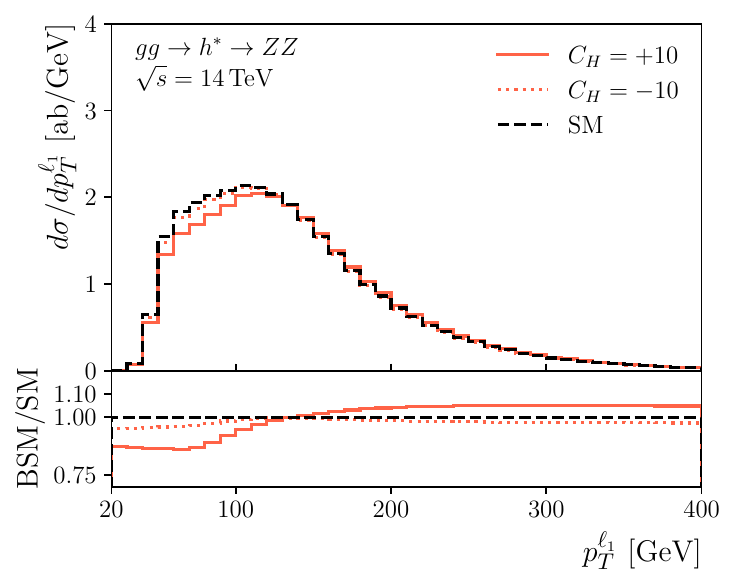}
\includegraphics[width=0.49\textwidth]{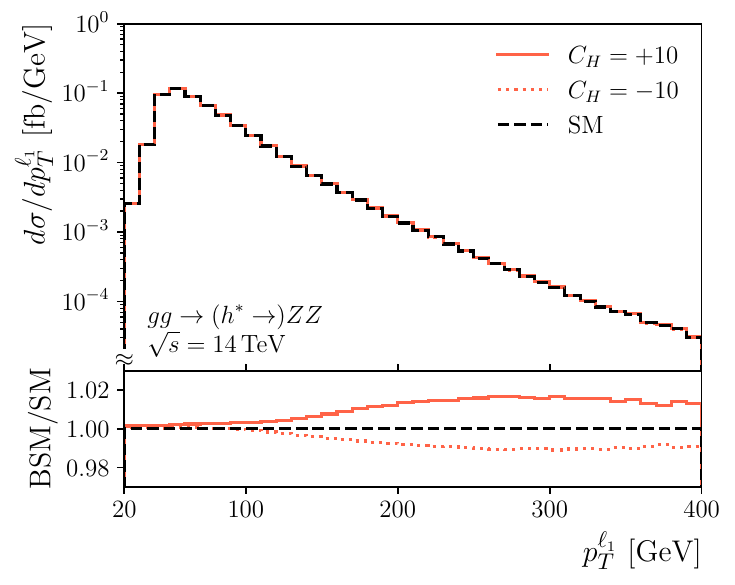}
\vspace{-2mm}
\caption{\label{fig:pTl1CH_10} Identical to~\cref{fig:m4lCH_10}, but displaying the transverse momentum of the leading lepton $p_T^{\ell_1}$.}
\end{figure}

In~\cref{sec:distributions}, we presented the $m_{ZZ}$ and $p_T^{\ell_1}$ distributions for $C_H = \pm 50$. In this appendix, we repeat the analysis for the smaller benchmark values $C_H = \pm 10$ to allow a direct comparison. The corresponding $m_{ZZ}$ and $p_T^{\ell_1}$ distributions are shown in Figures~\ref{fig:m4lCH_10} and~\ref{fig:pTl1CH_10}, respectively. As in the main text, the upper panels display the Higgs-channel contribution alone, while the lower panels include the combined effects of the Higgs channel, the continuum box background, and their interference.

In the upper panel of~\cref{fig:m4lCH_10}, the BSM prediction for $gg \to h^\ast \to ZZ \to 4\ell$ lies below the SM result for $275 \, {\rm GeV} \lesssim m_{ZZ} \lesssim 400 \, {\rm GeV}$. This behavior is driven by Higgs propagator corrections --- the only terms quadratic in~$C_H$ ---which reduce the real part of the amplitude, causing destructive interference and, for $C_H = +10$, a suppression of about $25\%$. Including the continuum box background and its interference turns this into a small positive effect below threshold. The $m_{ZZ}$ distributions also show modifications in the high-mass tail, most clearly for $C_H = +10$. Compared to~\cref{fig:m4lCH}, the overall effects in~\cref{fig:m4lCH_10} are smaller but remain qualitatively distinct from the changes induced by tree-level insertions of dimension-six SMEFT operators.

In~\cref{fig:pTl1CH_10}, we show the $p_T^{\ell_1}$ distributions for $C_H = \pm 10$, with the $C_H = \pm 50$ results given in the main text in~\cref{fig:pTl1CH}. The shape of the modifications is similar for both sets of $C_H$ values, with the main difference being the overall magnitude, which is drastically reduced for $C_H = \pm 10$ compared to $C_H = \pm 50$. Taken together, Figures~\ref{fig:m4lCH_10} and~\ref{fig:pTl1CH_10} indicate that for $C_H = \pm 10$ the BSM effects in the relevant $pp \to ZZ \to 4 \ell$ distributions are at the level of roughly $2\%$.

\section{Diagnostic test in case of $pp \to ZZ \to 2\ell2\nu$ channel}
\label{sec:llnunuDiagnostics}

\begin{figure}[!t]
\centering
\includegraphics[width=0.5\textwidth]{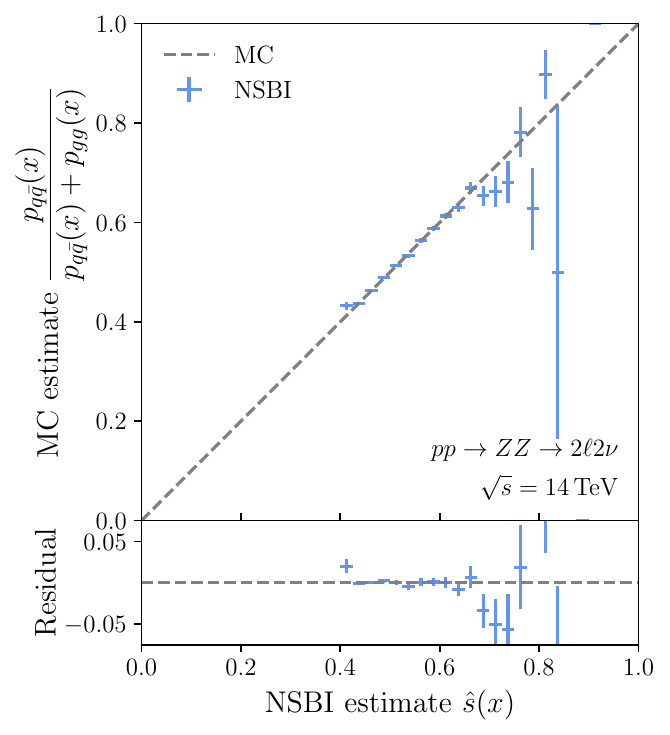}
\vspace{-2mm}
\caption{Calibration closure of the NN-estimated probability ratio, similar to Figure~\ref{fig:carl_calibration}, but for the $pp \to ZZ \to 2 \ell 2 \nu$ channel. The results displayed include events from the $gg \to (h^{\ast} \to) \, ZZ \to 2\ell 2\nu$, $q\bar{q} \to ZZ \to 2\ell 2\nu$, and $q\bar{q} \to W^{+}W^{-} \to 2\ell 2\nu$ channels.} \label{fig:carl_calibration_2l2v}
\end{figure}

In~\cref{sec:nn_calibration}, we carried out a calibration closure test of the NN-estimated probability ratio for the $pp \to ZZ \to 4\ell$ channel. A similar test is performed for the $pp \to ZZ \to 2\ell 2\nu$ channel in this appendix. For the probability ratio defined in~(\ref{eq:carl}), we compare the classifier output, $\hat{s}(x)$, to the fraction of events originating from the numerator hypothesis, using a balanced dataset, as shown in~\cref{fig:carl_calibration_2l2v}. Due to the partial reconstruction of the final-state kinematics, the classifier performance is reduced compared to the $pp \to ZZ \to 4\ell$ case, as evidenced by its output range being limited to approximately $0.4$ to $0.8$. Despite this, the calibration curve in the upper panel follows a near-diagonal trend within this range. The~corresponding NSBI results are consistent with this behavior, indicating good calibration within statistical uncertainties.

\end{document}